\documentclass[preprint,aps,eqsecnum]{revtex4}
\usepackage{graphicx}
\usepackage{bm}
\usepackage{amsmath}
\usepackage{amssymb}
\usepackage{amsfonts}
\usepackage{fancybox}
\usepackage[dvips]{color}
\newcommand{\half}{\frac{1}{2}}

\newcommand{\der}{\partial}

\newcommand{\Tr}{\mbox{\rm Tr}}
\newcommand{\dsl}{\partial\kern-0.55em\raise 0.14ex\hbox{/}}

\newcommand{\pd}[2]{\frac{\partial #1}{\partial #2}}

\newcommand{\bfk}{\bm{k}}

\newcommand{\bfp}{\bm{p}}

\newcommand{\bfu}{\bm{u}}

\newcommand{\bfP}{\bm{P}}

\begin{document}

\preprint{KYUSHU-HET-105}

\title{More about the Wilsonian analysis on the pionless NEFT}

\author{Koji Harada}
\email{harada@phys.kyushu-u.ac.jp}
\author{Hirofumi Kubo}
\email{kubo@higgs.phys.kyushu-u.ac.jp}
\author{Atsushi Ninomiya}
\email{ninomiya@higgs.phys.kyushu-u.ac.jp}
\affiliation{Department of Physics, Kyushu University\\ 
Fukuoka 812-8581 Japan}

\date{\today}

\begin{abstract}
 We extend our Wilsonian renormalization group (RG) analysis on the
 pionless nuclear effective theory (NEFT) in the two-nucleon sector in
 two ways; on the one hand, (1) we enlarge the space of operators up to
 including those of $\mathcal{O}(p^4)$ in the $S$ waves, and, on the
 other hand, (2) we consider the RG flows in higher partial waves ($P$
 and $D$ waves). In the larger space calculations, we find, in addition
 to nontrivial fixed points, two ``fixed lines'' and a ``fixed surface''
 which are related to marginal operators. In the higher partial wave
 calculations, we find similar phase structures to that of the $S$
 waves, but there are \textit{two} relevant directions in the $P$ waves
 at the nontrivial fixed points and \textit{three} in the $D$ waves. We
 explain the physical meaning of the $P$-wave phase structure by
 explicitly calculating the low-energy scattering amplitude. We also
 discuss the relation between the Legendre flow equation which we employ
 and the RG equation by Birse, McGovern, and Richardson, and possible
 implementation of Power Divergence Subtraction (PDS) in higher partial
 waves.
\end{abstract}


\maketitle

\section{Introduction}


Nuclear effective field theory
(NEFT)~\cite{Weinberg:1990rz,Weinberg:1991um,Weinberg:1992yk} is a
low-energy effective field theory of nucleons based on the general
principles of quantum field theory and the symmetries of the underlying
theory of hadrons, QCD.  (See Ref.~\cite{Beane:2000fx, Bedaque:2002mn,
Epelbaum:2005pn} for reviews.) At very low energies, where even the
pions are regarded as ``heavy,'' interactions of nonrelativistic
nucleons are simulated by (infinitely many) contact operators. Such a
theory is called \textit{pionless} NEFT with the physical cutoff scale
being around the pion mass, while the \textit{pionful} NEFT
is needed at higher energies, where the effects of the exchange of pions
are explicitly taken into account. Because NEFT contains infinitely many
operators, one needs an organizing principle, called power counting, to
systematically calculate physical quantities to a certain order.


It is interesting to note that the actual two-nucleon system in the $S$
waves is fine-tuned. The scattering lengths are unnaturally large
compared to the scale characteristic to the two-nucleon
interaction. From the RG point of view, this unnaturalness may be
rephrased as that the system is very close to the nontrivial fixed point
(or better, to the critical
surface)~\cite{Kaplan:1998tg,Kaplan:1998we}. The two-nucleon system is
thus nonperturbative due not only to the strong coupling, but also to
the closeness to the critical surface.


The unnaturalness of NEFT makes the power counting issue
complicated. The so-called Naive Dimensional Analysis
(NDA)~\cite{Manohar:1983md} works for perturbative systems, but does not
account for fine-tuning. One needs a power counting which encodes the
fine-tuning. There are a lot of papers~\cite{Kaplan:1996xu,
Kaplan:1998tg, Kaplan:1998we, Gegelia:1998iu, vanKolck:1998bw,
Fleming:1999ee, Beane:2001bc, Birse:2005um, Nogga:2005hy,
Epelbaum:2006pt} devoted to the power counting issues in NEFT, and it is
still an important subject of discussions.


In a previous paper~\cite{Harada:2006cw}, two of the present authors
performed a Wilsonian RG analysis~\cite{Wilson:1973jj, Wilson:1974mb,
Wegner:1972ih, Polchinski:1983gv} of pionless NEFT to determine the
power counting of the operators in the ${}^1S_0$ and ${}^3S_1$-${}^3D_1$
channels on the basis of the scaling dimensions at the nontrivial fixed
point. We employed the Legendre flow equation~\cite{Nicoll:1977hi,
Wetterich:1989xg, Bonini:1992vh, Morris:1993qb, Berges:2000ew} as a
formulation of nonperturbative RG, and we reproduced known results
obtained by Birse et al.~\cite{Birse:1998dk} up to $\mathcal{O}(p^2)$ in
a completely field theoretical fashion.  We also emphasized the phase
structure and identified the inverse of the scattering length as the
order parameter. In the \textit{strong coupling phase}, there is a
coupling which grows as the floating cutoff $\Lambda$ is
lowered, and there is a bound state, while in the \textit{weak coupling
phase}, all the flows run into the trivial fixed point and there is no
bound state. The determination of the power counting on the basis of the
scaling dimensions is a unifying principle of EFT, useful especially for
the cases in which nonperturbative dynamics is important.


One of the good features of Wilsonian RG analysis of EFT is that it
provides an overview of all the possible theories consistent with the
symmetries, not restricted only to the one that describes the real
world. In other words, it characterizes the physical system in a broad
perspective in terms of the RG flows.


Although the formulation of the Legendre flow equation does not contain
any approximation, in order to solve it, one needs an approximation; the
restriction of the space of operators. It is important to note that the
restriction does not require an \textit{a priori} power counting more
than just the canonical dimensional counting. The point is to include a
sufficient number of operators. The scaling dimensions are determined
for the linear combinations of the operators included at each fixed
point.


It is important to note that when the space of operators is enlarged,
the results (e.g., the scaling dimensions) are expected to always
converge to the true values. The question is \textit{how fast} they
converge. In order to see it, it is necessary to actually enlarge the
space of operators.  If the enlargement does not alter the results of
the calculations with smaller space very much, we may conclude that the
results are close to the true values. It has been argued that the small
dependence on the choice of cutoff functions also suggests fast
convergence~\cite{Litim:2002cf}.


Since our formulation is quite general, we can readily do a similar
analysis to other systems. For example, one can investigate higher
partial waves, where higher derivative terms play an important
role. Note however that, in general, higher partial waves are physically
not so significant at very low energies and that the physical
two-nucleon system does not seem to be fine-tuned in those higher
partial waves. Nevertheless, it is interesting to perform the RG
analysis for higher partial waves, because it provides a better
characterization of the physical two-nucleon systems than just that of
the angular momentum. (How many nontrivial fixed points there are? How
far is the physical system from them?) This kind of information may be
useful in the study of other systems described by a similar EFT.


One may think of the so-called ``halo nuclei'' as such an example. A
typical halo nucleus consists of a core (e.g., alpha-particle) and a few
``halo'' nucleons, and an EFT is proposed to describe such a
system~\cite{Bertulani:2002sz}.  A particularly interesting example is the
nucleon-alpha scattering in the $P$-wave, for which the $p_{3/2}$
channel displays a resonance just above the
threshold~\cite{Bertulani:2002sz, Bedaque:2003wa}. Such a resonance is
clearly related to a RG critical surface. The RG structure for 
nucleon-alpha scattering is very similar to that of the pionless NN
scattering, which we consider in this paper. We provide
an account on the power counting of Refs.~\cite{Bertulani:2002sz,
Bedaque:2003wa} on the basis of Wilsonian RG analysis


In this paper, we continue our previous study and perform a Wilsonian RG
analysis of pionless NEFT in two-nucleon sector. After recapitulating
the previous paper, we first enlarge the space of operators to include
those of $\mathcal{O}(p^4)$ in the ${}^1S_0$ channel. We find two
``fixed lines'' and a ``fixed surface'' besides trivial and nontrivial
fixed points. The nontrivial fixed point that we think the most relevant
to the real world persists, and the scaling dimensions of the
eigen-operators of the linearized RGE do not change, consistent with
Ref.~\cite{Birse:1998dk}.

Second, we consider the $P$ waves and obtain the phase structure by
solving the RG equations. The phase structure is similar to that in
the $S$ waves, but there are \textit{two} relevant operators in the
$P$ waves at the nontrivial fixed points.  We also discuss briefly the
$D$ waves and find that there are \textit{three} relevant ones.




The structure of the paper is the following: In Sec.~\ref{sec:review} we
recapitulate the main points of the previous paper~\cite{Harada:2006cw}
in order to introduce the notations and the main concepts in our
analysis. Since the present paper closely follows the discussions given
in Ref.~\cite{Harada:2006cw}, we expect that the readers intimately
consult with Ref.~\cite{Harada:2006cw}.  In Sec.~\ref{sec:p^4} we
discuss the enlargement of the space of operators. The calculations for
the higher partial waves are given in Sec.~\ref{sec:hpw}. We also
discuss possible extensions of Power Divergence Subtraction (PDS)
renormalization for higher partial waves. The summary and discussions
are given in Sec.~\ref{sec:summary}. In Appendix~\ref{sec:equiv}, we
discuss the relation between the Legendre flow equation in the sharp
cutoff limit and the RG equation used by Birse et
al.~\cite{Birse:1998dk}. The cutoff function dependence of the results
is studied in Appendix~\ref{sec:n-dep}.

\section{Nonperturbative RG analysis for the two-nucleon system at very low
 energies}
\label{sec:review}

\subsection{Power counting and the scaling dimensions}

The most basic idea behind the power counting is the order of magnitude
estimate based on dimensional analysis. For the two-nucleon system at
very low energies described by the pionless NEFT, there is a physical
cutoff scale $\Lambda_0 \sim \mathcal{O}(m_\pi)$, above which the
effective theory is not applicable.  Since it is the only scale (except
for the nucleon mass, $M$, see the footnote 3 of
Ref.~\cite{Harada:2006cw} and the section 1.2.1 of
Ref.~\cite{Kaplan:2005es}) which appears in the pionless NEFT, the
dimensional analysis for the pionless NEFT is based on this scale.

Quantum fluctuations may change the classical dimensional analysis. The
quantum counter part of the dimension is the \textit{scaling
dimension}, which can be obtained by the RG analysis. It is therefore
natural to consider the power counting based on the scaling dimensions.
Wilsonian, or nonperturbative, RG is a suitable tool to handle the
quantum fluctuations.

It is known that there is another scale in the two-nucleon system, the
scattering length in an $S$ wave, the inverse of which is much smaller
than $\Lambda_0$. It can be understood that such a scale is not
fundamental, but is a result of the fine-tuning of the parameters of the
EFT.  In the RG language, the fine-tuning is closely related to the
existence of a nontrivial fixed point and a critical surface of the
RGE. It has been shown that on the critical surface the scattering
length is infinite. In order to get a large scattering length, the
coupling constants must be fine-tuned to be near the critical surface.

Around the nontrivial fixed points the scaling dimensions are
drastically different from their classical values, the canonical
dimensions. The coupling which corresponds to the scattering length
classically has negative dimension, indicating that it gets less
important at lower energies. But near the nontrivial fixed point (or the
critical surface), the (quantum) scaling dimension is positive, so the
corresponding interaction becomes more important at lower energies. On
the other hand, near the trivial fixed point, where all the coupling
constants are small, the scaling dimensions are the same as the
classical values. There the quantum fluctuations do not alter the
importance of the interaction.

A relevant operator, which has a positive scaling dimension, controls
the deviation of the coupling constants from a critical surface. In
order to get a scale much smaller than $\Lambda_0$, the relevant
operator must be fine-tuned. Near a nontrivial fixed point, the number
of the parameters to be fine-tuned is that of the relevant operators.


In determining the \textit{power counting}, one usually does two things:
(1) choose a set of operators one is going to work with, and (2)
determine the importance of the operators. In a conventional power
counting scheme, one first determines the importance of the operators
(2) by assuming some scaling property, and it then leads to the set of
operators (1) to a given order. On the other hand, in the Wilsonian RG
approach we only choose a set of operators (1). The importance of the
operators (2) is determined as an output by the scaling dimensions of
them.

In most cases, it is unlikely to miss important operators in both
approaches, as far as one includes all the operators of low canonical
dimensions to a certain order. What matters is the determination of
the importance of the operators (the scaling property).


Up to the point where the fixed points and the scaling dimensions are
obtained in the Wilsonian RG approach, however, no contact has been made
with where the real world is in the RG flow. The Wilsonian RG approach
reveals all the possible theories described by the same action without
the knowledge of the actual systems. It is the strength as well as the
weakness of the approach. Namely, one can determine possible scaling
properties consistent with the action (the symmetry, spacetime
dimensions, and the degrees of freedom) while one still needs additional
information about the system (e.g., a large scattering length) one is
trying to describe, in order to determine where it is in the RG flow.

Convergence of the approximation has different meanings in the
conventional and the Wilsonian RG approaches: in the conventional
approach, the validity of the power counting is examined (in most cases,
numerically) \textit{whether} the expansion converges. With a wrong
assumption of scaling property, the expansion fails to converge, i.e.,
even though one includes more operators, the results (the phase shift of
a specific channel, for example) do not improve. See
Ref.~\cite{Fleming:1999ee} for the important example. On the other hand,
in the Wilsonian RG approach, the results (the scaling dimensions at a
specific fixed point, for example) improve as the space of the operators
is enlarged. The question is \textit{how fast} the results converge.


\subsection{Legendre flow equation}

Legendre flow equation~\cite{Nicoll:1977hi, Wetterich:1989xg,
Bonini:1992vh, Morris:1993qb, Berges:2000ew} is one of the
implementations of Wilsonian RGE. It is formulated as a RGE for the
infrared (IR) cutoff effective action $\Gamma_{\Lambda}[\Phi]$ called
\textit{effective averaged action}, in which the quantum fluctuations
above the cutoff have been integrated. The averaged action is the
generator of the one-particle irreducible (1PI) vertex functions
containing only the fluctuations $p \agt \Lambda$.

The Legendre flow equation is given by
\begin{equation}
 \frac{d\Gamma_\Lambda}{d\Lambda}=\frac{i}{2}\Tr
  \left[
   \frac{dR_\Lambda}{d\Lambda}\left(\Gamma_{(2)}+R_\Lambda\right)^{-1}
  \right],
\end{equation}
where $\Gamma_{(2)}$ stands for the second derivative of the averaged
action $\Gamma_\Lambda$ with respect to the fields, and is the inverse
of the full propagator dressed by the vertices with classical field
insertions, containing only the fluctuations $p \agt \Lambda$, and the
$\Tr$ denotes the integration over momentum and also the trace in the
internal space. The function $R_{\Lambda}$ effectively cuts off the IR
part of the fluctuations. Our choice is, as in the previous paper, as
follows;
\begin{equation}
 {R}_{\Lambda}(\bfp^2)=
\frac{\bfp^2}{2M}
\left[1-\exp\left[\left(\frac{\bfp^2}
{{\Lambda}^2}\right)^n\right]\right]^{-1},
\label{cutoffR}
\end{equation}
where $\bfp$ is the three-momentum.  In the $n\rightarrow \infty$ limit, it
becomes a sharp cutoff. Another choice is adopted in
Ref.~\cite{Birse:2008wt}. We derive the RGE for an arbitrary value of
$n$, but the results look so complicated that we present only the
results in the $n\rightarrow \infty$ limit. The $n$-dependence is
studied in Appendix~\ref{sec:n-dep}.

In the application to the two-nucleon system, we have drastic
simplification of the RGE. (i) Because of the nonrelativistic feature of
the system, we do not include anti-particles. Thus there are no
self-energy corrections, nor tadpole contributions.  (ii) In the
two-nucleon sector only the four-nucleon (4N) operators contribute.
From these, we end up with the one-loop diagrams involving two vertices
with the tree-level propagators contributing to the Legendre flow
equation. The multi-loop effects are encoded in the cutoff dependence of
the (infinitely many) coupling constants.

In Appendix~\ref{sec:equiv}, we discuss the relation between
the Legendre flow equation in the sharp cutoff limit and the RG equation
used by Birse et al. for the potential $V$,
\begin{equation}
 \pd{V}{\Lambda}=\frac{M}{2\pi^2}V(k', \Lambda, p ;\Lambda)
  \frac{\Lambda^2}{\Lambda^2-p^2}V(\Lambda, k, p ;\Lambda).
  \label{birse_rge}
\end{equation}
 (See Ref.~\cite{Birse:1998dk} for details.) It is shown that they are
essentially equivalent.  


As we emphasized in Ref.~\cite{Harada:2006cw}, there is no obvious way
of imposing a Galilean invariant cutoff at the averaged action
level~\cite{Birse:2008wt}, so that the inclusion of the IR cutoff
function in the averaged action should be understood as a symbolic
one. Note that the IR cutoff function constrains the momenta of
individual particles, but they are not invariants under Galilean
transformations. Fortunately, however, the correct way of implementing a
cutoff is clear in the two-nucleon system; to impose the cutoff on the
relative three momentum.  See Sec.~4.1 in Ref.~\cite{Harada:2006cw} for
the explicit manipulation. Note that it is \textit{not} an
approximation, but the \textit{only} way that we know to implement the
cutoff in a Galilean invariant way.

\subsection{Pionless NEFT up to $\mathcal{O}(p^2)$ in the ${}^1S_0$
 channel}
\label{sec:p2}

Although the Legendre flow equation is \textit{exact}, one needs an
approximation to solve it. We consider a simple truncation of the space
of operators. We retain only the operators with derivatives up to a
certain order. We simply count the number of spatial derivatives
($\nabla \sim p$) and a time derivative is counted as two spatial
derivatives ($\der_t \sim p^2$).  The approximation is based on our hope
that, even though some operators get large anomalous dimensions, their
``ordering'' of importance would not change very much; the lower the
canonical dimension is, the lower the scaling dimension would be. We
also expect that there are mixings among operators. In
Ref.~\cite{Harada:2006cw}, we consider the following ansatz for the
averaged action up to $\mathcal{O}(p^2)$,
\begin{eqnarray}
\Gamma_{\Lambda}^{(p^2)}&=&\int d^4x
 \Bigg[
 N^{\dagger}\left(i\partial_t+\frac{{\nabla}^2}{2M}\right)N
    -C_0  \left(N^TP^{({}^1S_0)}_aN\right)^\dagger
 \left(N^TP^{({}^1S_0)}_aN\right) 
 \nonumber \\
 &&{}+C_2
 \left[
 \left(N^TP^{({}^1S_0)}_aN\right)^\dagger
 \left(N^TP^{({}^1S_0)}_a\overleftrightarrow{\nabla}^2 N\right)
 + \mbox{h.c.}
 \right] \nonumber \\
 &&{}+2B
 \left[
 \left\{
 N^TP^{({}^1S_0)}_a \left(i\der_t+\frac{\nabla^2}{2M}\right) N
 \right\}^\dagger
 \left(N^TP^{({}^1S_0)}_aN\right)
 + \mbox{h.c.} 
 \right] 
 \Bigg],
 \label{second}
\end{eqnarray}
for the ${}^1S_0$ channel, where $P^{({}^1S_0)}_a$ is the projection
operator to the channel,
\begin{equation}
 P^{({}^1S_0)}_a=\frac{1}{\sqrt{8}}\sigma^2\tau^2\tau^a, 
  \label{proj1s0}
\end{equation}
and $\overleftrightarrow{\nabla}^2 \equiv \overleftarrow{\nabla^2} +
\overrightarrow{\nabla^2} - 2\overleftarrow{\nabla} \cdot
\overrightarrow{\nabla}$.  As we emphasized in a previous
paper~\cite{Harada:2005tw}, it is important to include the so-called
``redundant operators,'' which can be eliminated from the action by
using the equations of motion.

By introducing the following dimensionless coupling constants,
\begin{equation}
 x=\frac{M\Lambda}{2\pi^2}C_0, \quad y=\frac{M\Lambda^3}{2\pi^2}4C_2,
  \quad z=\frac{\Lambda^3}{2\pi^2}B, 
  \label{xyz}
\end{equation}
the RGEs can be expressed as
\begin{eqnarray}
 \frac{dx}{dt}&=&
  -x
  -\Biggl[
  x^2+2xy+y^2+2xz+2yz+z^2
  \Biggr], \label{xbeta}\\
 \frac{dy}{dt}&=&
  -3y
  -\Biggl[
  \frac{1}{2}x^2+2xy+\frac{3}{2}y^2+yz-\frac{1}{2}z^2
  \Biggr], \label{ybeta}\\
 \frac{dz}{dt}&=&
  -3z
  +\Biggl[
  \frac{1}{2}x^2+xy+\half y^2-xz-yz-\frac{3}{2}z^2
  \Biggr], \label{zbeta}
\end{eqnarray}
where $t=\ln\left(\Lambda_0/\Lambda\right)$. We found a nontrivial fixed
point, 
\begin{equation}
 (x^\star, y^\star, z^\star) = \left(-1, -\half, \half\right),
\label{AofRef18}
\end{equation}
as well as the trivial one $(0,0,0)$. The eigenvalues and the
corresponding eigenvectors of the linearized RGE at the nontrivial fixed
points are found to be
\begin{align}
 &\nu_1=+1: \quad \bfu_1=
 \left(
 1,1,-1
 \right), \\
 &\nu_2=-1: \quad \bfu_2=
 \left(
 0,-1,1
 \right), \\
 &\nu_3=-2: \quad \bfu_3=
 \left(
 2,-1, -2
 \right).
\end{align}
(Note that the signs and the normalizations of the eigenvectors are
arbitrary.) The eigenvector associated with the positive eigenvalue
$\nu_1$ corresponds to the scattering length.

In Ref.~\cite{Harada:2005tw} we performed a similar RG analysis based on
the cutoff independence of the amplitude and found the same scaling
dimensions with slightly different eigenvectors. Since eigenvectors are
not universal quantities and the approximations are different, it is
therefore not surprising that the eigenvectors are different. The
important thing is that the eigenvalues, which are considered as the
universal ones, actually agree.  The fact that the first two
eigenvectors agree reflects that these two approximations are similar.

These results are obtained with a severe restriction of the space of
operators.  One should examine the validity of the approximation by
actually enlarging the space of operators. It is expected that some
properties, i.e., the scaling dimensions at the nontrivial fixed points,
are universal and would not change very much under the enlargement if
the approximation is good, while the others such as the directions of
eigenvectors may change. It is well known that the restriction of the
space of operators causes various artefacts~\cite{Margaritis:1987hv,
Morris:1994ki, Morris:1999ba}.

In the next Section, we investigate the effects of the enlargement of
the space of operators and compare the results with the previous ones
presented in this section.

\section{Pionless NEFT up to $\mathcal{O}(p^4)$ in the ${}^1S_0$ channel}
\label{sec:p^4}

\subsection{Independent operators}


In this section, we enlarge the space of operators in the ${}^1S_0$
channel up to including those of $\mathcal{O}(p^4)$. (The results for
the ${}^3S_1$-${}^3D_1$ channel are essentially the same as those for
the ${}^1S_0$, but a bit more complicated.)  We consider the following
ansatz for the averaged action,
\begin{align}
 \Gamma_{\Lambda}^{(p^4)}= \Gamma_{\Lambda}^{(p^2)}  \qquad &\nonumber \\
  +
\int d^4x \Bigg[&
 -C_{41}
 \left[
 \left(N^TP^{({}^1S_0)}_aN\right)^\dagger
 \left(N^TP^{({}^1S_0)}_a\overleftrightarrow{\nabla}^4 N\right)
 + \mbox{h.c.}
 \right] 
 \nonumber \\
 & -C_{42}
 \left(N^TP^{({}^1S_0)}_a\overleftrightarrow{\nabla}^2 N\right)^\dagger
 \left(N^TP^{({}^1S_0)}_a\overleftrightarrow{\nabla}^2 N\right)
 \nonumber \\
 & + 2B_{1}
 \left[
 \left(N^TP^{({}^1S_0)}_aN\right)^\dagger
 \left\{
 N^TP^{({}^1S_0)}_a \left(i\der_t+\frac{\nabla^2}{2M}\right)^2 N
 \right\}
 + \mbox{h.c.}
 \right]
 \nonumber \\
 & +4B_{2}
 \left\{
 N^TP^{({}^1S_0)}_a \left(i\der_t+\frac{\nabla^2}{2M}\right) N
 \right\}^\dagger
 \left\{
 N^TP^{({}^1S_0)}_a \left(i\der_t+\frac{\nabla^2}{2M}\right) N
 \right\}
 \nonumber \\
 & -2B_{3}
 \left[
 \left(N^TP^{({}^1S_0)}_a\overleftrightarrow{\nabla}^2 N\right)^\dagger
 \left\{
 N^TP^{({}^1S_0)}_a \left(i\der_t+\frac{\nabla^2}{2M}\right) N
 \right\}
 + \mbox{h.c.}
 \right]
 \Bigg].
 \label{average4}
\end{align}


We emphasize that the ansatz given above contains all the independent
operators consistent with Galilean invariance, parity, spin and isospin
invariance to the given order. There are actually two other operators
which satisfies these requirements,
\begin{align}
 &\left[
 \left(N^TP^{({}^1S_0)}_aN\right)^\dagger
 \left\{
 N^TP^{({}^1S_0)}_a \overleftrightarrow{\nabla}^2
 \left(i\der_t+\frac{\nabla^2}{2M}\right) N
 \right\}
 + \mbox{h.c.}
 \right], \\
 &\left[
 \left(N^TP^{({}^1S_0)}_aN\right)^\dagger
 \left\{
 N^TP^{({}^1S_0)}_a 
 \left(i\overleftarrow{\der_t}+\frac{\overleftarrow{\nabla}^2}{2M}\right)
 \left(i\der_t+\frac{\nabla^2}{2M}\right) N
 \right\}
 + \mbox{h.c.}
 \right],
\end{align}
but they are written as linear combinations of the operators contained
in the averaged action up to total derivatives, and thus we dropped them.
The relations may be expressed most clearly in momentum space. Using
momentum conservation, $(p_1+p_2)^\mu=(p_3+p_4)^\mu$, we obtain
\begin{align}
 r_{12}\left(S_1+S_2\right)+r_{34}\left(S_3+S_4\right)
 =& r_{12}\left(S_1+S_4\right)+r_{34}\left(S_1+S_2\right) \nonumber \\
 &{}-
 \frac{1}{4M}\left(r_{12}^2+r_{34}^2\right)
 +\frac{1}{2M}r_{12}r_{34}, \\
 S_1S_2+S_3S_4 
 =& \left(S_1+S_2\right)\left(S_3+S_4\right)
 -\half \sum_{i=1}^4S_i \nonumber \\
 &{}
 +\frac{1}{32M}\left(r_{12}^2+r_{34}^2\right)
 -\frac{1}{16M}r_{12}r_{34},
\end{align}
where we have introduced the notations,
\begin{equation}
 r_{ij}\equiv \left(\bfp_i-\bfp_j\right)^2,\quad
  S_i\equiv p^0_i-\frac{\bfp_i^2}{2M}.
  \label{def:r:S}
\end{equation}

It is also important to note that we have included the interaction
($B_1$) which depends not only on the total energy of the two nucleons,
but on the individual energies. The potential corresponding to this
interaction is not considered in Ref.~\cite{Birse:1998dk}.

We do not include the relativistic correction terms such as $N^\dagger
\left(\nabla^4/8M^3\right) N$ in the averaged action because it breaks
Galilean invariance. It means that we are considering the system at very
low energies so that the relativistic corrections can be neglected. The
effect of this term is estimated to be smaller than any of the terms we
have included in Eq.~(\ref{average4}) due to the additional
$\mathcal{O}\left((\Lambda/M)^2\right)$ suppression.

\subsection{RG equations}
\label{sec:rge}


The Legendre flow equation may be obtained in a similar manner as in
Ref.~\cite{Harada:2006cw}, which generalizes the set of RGEs
(\ref{xbeta}), (\ref{ybeta}), and (\ref{zbeta}).


By introducing the following dimensionless coupling constants,
\begin{equation}
 u_i=\frac{M\Lambda^5}{2\pi^2}16C_{4i} \ (i=1,2), \quad
 z_i=\frac{\Lambda^5}{2\pi^2 M}B_i \ (i=1,2), \quad
 z_3=\frac{\Lambda^5}{2\pi^2}4B_3,
 \label{dimless}
\end{equation}
together with $x$, $y$, and $z$ defined in (\ref{xyz}),
we have a set of the RGEs for these variables in the $n\rightarrow
\infty$ limit,
\begin{align}
 \frac{dx}{dt}=&-x
 -x(x+2y+2z+2u_1-2z_1)
 -y(y+2z+2u_1-2z_1) \nonumber \\
 &
 -z(z+2u_1-2z_1)
 -u_1(u_1-2z_1)
 -z_1^2, \\
 \frac{dy}{dt}=&-3y
 -x\left(\half x+2y+u_1+u_2+z_1+z_3\right)
 -y\left(\frac{3}{2}y+z+2u_1+u_2+z_3\right) \nonumber \\
 &
 +z\left(\half z-u_2-2z_1-z_3\right)
 -u_1\left(\half u_1+u_2+z_1+z_3\right)
 +u_2z_1 \nonumber \\
 &
 +z_1\left(\frac{3}{2}z_1+z_3\right), \\
 \frac{dz}{dt}=&-3z
 +x\left(\half x+y-z+u_1+z_1+z_2-z_3\right)
 +y\left(\half y-z+u_1+z_1+z_2-z_3\right) \nonumber \\
 &
 -z\left(\frac{3}{2}z+u_1-3z_1-z_2+z_3\right)
 +u_1\left(\half u_1+z_1+z_2-z_3\right) \nonumber \\
 &
 -z_1\left(\frac{3}{2}z_1+z_2-z_3\right), \\
 \frac{du_1}{dt}=&-5u_1
 -u_1(x+y+z+u_1-z_1), \\
 \frac{du_2}{dt}=&-5u_2
 -x(x+4y+2u_1+2u_2)
 -y(4y+4u_1+4u_2+2z_1+2z_3) 
 +z(2z_1+2z_3) \nonumber \\
 &
 -u_1(u_1+2u_2)
 -u_2(u_2+2z_1+2z_3)
 -z_1(3z_1+4z_3)
 -z_3^2, \\
 \frac{dz_1}{dt}=&-5z_1
 -z_1(x+y+z+u_1-z_1), \\
 \frac{dz_2}{dt}=&-5z_2
 +x(x+2y-2z+2u_1-2z_3)
 +y(y-2z+2u_1-2z_3) \nonumber \\
 &
 +z(z-2u_1-4z_1-4z_2+2z_3) 
 +u_1(u_1-2z_3)
 +z_1(3z_1+4z_2-2z_3) \nonumber \\
 &
 +z_2(z_2-2z_3)
 +z_3^2, \\
 \frac{dz_3}{dt}=&-5z_3
 +x(x+3y-z+2u_1+u_2-z_3)
 +y(2y-2z+3u_1+u_2+z_1+z_2-2z_3) \nonumber \\
 &
 -z(u_1+u_2+3z_1+z_2+2z_3)
 +u_1(u_1+u_2-z_3)
 +u_2(z_1+z_2-z_3) \nonumber \\
 &
 +z_1(3z_1+2z_2+z_3)
 +z_2z_3
 -z_3^2.
\end{align}

By setting $u_1=u_2=z_1=z_2=z_3=0$ by hand in the first three of these
RGEs and ignoring the rest, they reduce to the ones, (\ref{xbeta}),
(\ref{ybeta}), and (\ref{zbeta}), as they should. This is exactly what
we did in the $\mathcal{O}(p^2)$ calculations.


To analyze the RG flows, it is useful to determine the fixed points,
at which the coupling constants do not run. We use \textsl{Mathematica}
to solve the fixed point equations and found the following solutions.

\begin{enumerate}
 \item [A] $(0,0,0,0,0,0,0,0)$,
 \item [B]$\left(
       -1,-\frac{1}{2},\frac{1}{2},0,-\frac{4}{3},0,\frac{4}{3},\frac{4}{3}
       \right)$,
 \item [C] $\left(
       -\frac{9}{4},\frac{123}{44},\frac{21}{22},0,
       -\frac{92889}{15488},0,\frac{13689}{15488},-\frac{33831}{15488} 
       \right)$,
 \item [D] $\left(-25,\frac{175}{18}, -\frac{25}{6}, u_1, 20-2u_1,
       \frac{-130+9u_1}{9}, \frac{260-18u_1}{9}, \frac{10}{3}
       \right)$,
 \item [E] $\left(
       0,0,0,0,u_2,0,X_{\pm},
       \frac{-5 X_{\pm} + X_{\pm}^2 + 5 u_2 + 2 X_{\pm} u_2 + {u_2}^2}{10}
       \right)$,
 \item [F] $\left(\!-25,y,-y,u_1,
       \frac{23125 - 1000 u_1 - 1700 y + 36 y^2}{500},
       \!-20\!+\!u_1,\frac{19375 - 1000 u_1 + 300 y - 36y^2}{500},
       \frac{-1875 + 1000 y - 36 y^2}{500}\right)$,
\end{enumerate}
where we have introduced 
\begin{equation}
 X_{\pm}= 5 \pm 2{\sqrt{5}}{\sqrt{-u_2}} - u_2.
\end{equation}


The first fixed point [A] is the trivial fixed point.  The second one
[B] corresponds to the fixed point we considered in the previous paper
(the fixed point Eq.~(\ref{AofRef18})) because their first three
coordinates coincide with each other. Note that, since the restriction
of the space is nothing but the projection to the lower dimensional
subspace, the fixed point Eq.~(\ref{AofRef18}) has the components which
we naturally expect the corresponding fixed point to have if the
approximation is good. As we explained in Ref.~\cite{Harada:2006cw},
this fixed point is the most relevant to the real two-nucleon systems,
and we will discuss it later in detail.  The third one [C] is a strange
fixed point. We calculate the scaling dimensions at this point, and
found that they are: $-8, -13/2, -13/2, 5, 3, 1$, and $(-5\pm
i\sqrt{41})/2$. Because of the appearance of the complex scaling
dimensions, we suspect that this is an artefact of the truncation. We
will give the argument shortly.


The fourth [D] and the fifth [E] are actually ``fixed lines.'' The [D]
is a fixed point for an arbitrary $u_1$, and the [E] for an arbitrary
$u_2\le 0$. The [D] is a straight line, while the [E] is a parabolic
curve with the peak at $(0,0,0,0,0,0,5,0)$.  The sixth [F] is an even
stranger ``fixed surface.'' Interestingly the scaling dimensions are the
same on the whole fixed lines (surface). The scaling dimensions at [D]
are: $5, 0, (11 \pm i\sqrt{239})/6$, $-7.86665\pm 3.58098 i$, and
$-1.35557 \pm 0.309653 i$. Since this also has complex scaling
dimensions, we suspect that it is an artefact too. See below. The [E] has
the scaling dimensions, $-5,-5,-5, -3, -1, 0, 2, 5$. Note that all the
scaling dimensions are integers, and the existence of a marginal
direction. Actually, a fixed point on it has been noticed by
Birse~\footnote{M.~C.~Birse, private communication. We thank him for
explaining his way of getting this fixed point and the scaling
dimensions of the perturbations around it.}. The [F] has the scaling
dimensions, $-5, (-11\pm \sqrt{41})/2, 0, 0, 2, 5, 7$. Note that there
are two marginal directions, which are tangential to the surface.


We consider the fixed point [C] and and the fixed line [D], which have
complex scaling dimensions, artefacts of the truncation. In the
following, we argue for this statement by giving a list of
observations. We admit that each of them is not completely convincing,
but they all suggest that fixed objects with complex scaling dimensions
are artefacts.  Of course, there is no definite criterion by which we
can decide whether a fixed object is an artefact.

The observations are the followings: (1) It is well known that the
truncation generates complex scaling dimensions~\cite{Morris:1994ki},
and it seems a general feature of truncation.  So the appearance of
complex scaling dimensions can be seen as a signal of the artefact of
the truncation and does not seem to have real physical relevance, at
least in most cases.
(2) In the literature, it is extremely rare for complex scaling
dimensions that do not seem to be artefacts to appear in physical
systems. To our best knowledge, only hierarchical Ising
models~\cite{PhysRevLett.48.767} and gravitational
collapse~\cite{Koike:1995jm, Hara:1996mc} are such systems. The former
are frustrated at every length scale, while the latter is known to have
critical limit cycle (discrete self-similarity).  We see that these
examples are very different from the simple system of self-interacting
nonrelativistic fermions.
(3) In a previous paper~\cite{Harada:2006cw} we found a fixed point with
complex scaling dimensions, but when enlarging the space of operators in
the present paper, we do not find the corresponding fixed point. It
disappears. This is a concrete example of a nontrivial fixed point which
appears with complex scaling dimensions in smaller space of operators,
and is in fact an artefact of the truncation. It is important to note
that the very existence of a nontrivial fixed point can be an artefact.
(4) It seems unlikely to get complex scaling dimensions by the method
employed by Birse \textit{et al.}~\cite{Birse:1998dk}, which is proved
in Appendix~\ref{sec:equiv} to be essentially equivalent to ours.  Note
that, even though the RGE for the potential to be solved is shown to be
the same for the both approaches, the way of solving it is different.
In the approach by Birse \textit{et al.} the scaling dimensions are the
sums of the powers of the monomial perturbation around the (inverse of
the) fixed point potential that is not expanded in powers of the energy.
The scaling dimensions obtained in such a way should be integers.  (5)
Because of the nonrelativistic feature of the present theory, only power
divergences are involved. It is therefore natural to expect integer
scaling dimensions. From this point of view, the fractional scaling
dimensions of the fixed surface [F] also seem to be artefacts.

Again, we cannot completely exclude the logical possibility that the
fixed objects persist in further enlargement of space of
operators, while their scaling dimensions become real.

Even though their physical relevance is unclear at this moment, it is
interesting to find such possibilities of several kinds of fixed objects
appear in the system. These possibilities have never been revealed in
other papers.  Compare with the work by Birse et
al.~\cite{Birse:1998dk}, for example. They started with an ansatz for a
fixed potential that depends only on the total energy, found the
solution, and analyzed perturbations around it. In our analysis, on the
other hand, we scanned the whole (though restricted) theory space and
found all possible fixed objects.


In any case, irrespective to whether these fixed objects are artefacts
or not, the most relevant fixed point to the realistic two-nucleon
systems is the fixed point [B]. It is easy to see that our solution is
the same as the one obtained by Birse et al.~\cite{Birse:1998dk} up to
including $\mathcal{O}(p^4)$. 


At the nontrivial fixed point [B], we find the following scaling
dimensions and corresponding eigenvectors,
\begin{align}
 &\nu_1=+1: \quad \bfu_1=
 \left(
 1,1,-1,0,\frac{11}{3},0,-\frac{11}{3},-\frac{11}{3} 
 \right), \\
 &\nu_2=-1: \quad \bfu_2=
 \left(
 0,-1,1,0,-4,0,4,4
 \right), \\
 &\nu_3=-2: \quad \bfu_3=
 \left(
 1,1,-\frac{5}{2},0,\frac{23}{3},0,-\frac{32}{3},-\frac{55}{6} 
 \right), \\
 &\nu_4=-3: \quad \bfu_4=
 \left(
 0,0,0,0,1,0,-1,-1
 \right), \\
 &\nu_5=-4: \quad \bfu_5=
 \left(
 1,-1,1,\frac{1}{2},0,3,0,0
 \right), \\
 &\nu_6=-4: \quad \bfu_6=
 \left(
 1,-1,1,-\frac{5}{2},6,0,6,0
 \right), \\
 &\nu_7=-4: \quad \bfu_7=
 \left(
 -2,-1,1,5,-6,0,0,3
 \right), \\
 &\nu_8=-5: \quad \bfu_8=
 \left(
 -\frac{1}{2},-\frac{1}{2},2,0,\frac{1}{6},0,\frac{22}{3},\frac{4}{3}
 \right).
\end{align}
Note that (i) all the scaling dimensions are integers. (ii) The first
three eigenvalues coincide with those obtained in the previous paper,
which justifies the identification of the fixed point [B] in the present
paper with the fixed point Eq.~(\ref{AofRef18}) in the previous
paper. (iii) The results are insensitive to the choice of the cutoff
function. We give the analysis of the $n$-dependence in
Appendix~\ref{sec:n-dep}.  (iv) Only the eigenvector $\bfu_5$ contains
the $z_1$ component, i.e., depends on individual energies of two
nucleons. This eigenvector is not considered in
Ref.~\cite{Birse:1998dk}.


In this section, the enlargement of the space of operators in the
two-nucleon system in the ${}^1S_0$ channel is considered. We found a
new kind of extended fixed objects which have (a) marginal
direction(s). We then concentrated on the fixed point which is the most
relevant to the realistic two-nucleon system, and found that the
universal properties we got in the previous paper are unchanged under
the enlargement of the space of operators, confirming that the
approximation is reliable.

\section{Higher partial waves}
\label{sec:hpw}


Operators which contribute in higher partial waves contain more
derivatives, and thus are less important at low energies. In a few
instances, however, higher partial waves contain some interesting
information. A well known example is the $p_{3/2}$ wave of the
$n$-$\alpha$ system, where a narrow resonance state exists near the
threshold, which is discussed in the NEFT context in
Refs.~\cite{Bertulani:2002sz,Bedaque:2003wa}. An important feature of
this system is that there are \textit{two} coupling constants to be
fine-tuned. In the following, we will show that it comes out very
naturally from our Wilsonian RG analysis. A similar analysis can be done
for the $D$ waves, and we show that, if there is a bound (or a
resonance) state near the threshold, there are also \textit{three}
couplings to be fine-tuned.

\subsection{$P$ waves}
\label{sec:pwave}


In the NN system there are four channels in the $P$ waves: ${}^1P_1$,
${}^3P_0$, ${}^3P_1$, and ${}^3P_2$-${}^3F_2$, where we consider the
mixing with an $F$ wave for the $J=2$ channel. For the ${}^1P_1$
channel, the interaction terms for the $\Gamma_\Lambda$ may be written
as
\begin{align}
 \Gamma_\Lambda^{int}=&\int d^4x\Bigg[
 - C_2^{(^1P_1)}\left(N^{T}P_i^{(^1P_1)}N\right)^{\dagger}
 \left(N^{T}P_i^{(^1P_1)}N\right) \nonumber \\
 &+ C_4^{(^1P_1)}\left\{
 \left(N^{T}P_i^{(^1P_1)}N\right)^{\dagger}
 \left(N^{T}P_i^{(^1P_1)}\overleftrightarrow{\nabla}^2N\right)
    +h.c. \right\}\nonumber\\
 &+ 2B_4^{(^1P_1)}\left[
 \left\{N^{T}P_i^{(^1P_1)}
 \left(i\partial_{t}+\frac{{\nabla}^2}{2M}\right)
 N\right\}^\dagger
 \left(N^{T}P_i^{(^1P_1)}N\right)
+ h.c.\right]
 \Bigg]
 \label{pwavegamma}
\end{align}
up to including $\mathcal{O}(p^4)$, where $P_i^{(^1P_1)}$ is the
projection operator to the ${}^1P_1$ channel, defined as
\begin{equation}
  P^{(^1P_1)}_i =
  \frac{\sqrt{3}}{4\sqrt{2}}\overleftrightarrow{\nabla}_i(i\sigma_2)(i\tau_2)
. 
\end{equation}
Similarly, one can define the other projection operators,
\begin{align}
 P^{(^3P_0)}_a=&
  \frac{1}{4\sqrt{2}}\overleftrightarrow{\nabla}_j
 (i\sigma_2\sigma_j)(i\tau_2\tau_a)
, \\
 P^{(^3P_1)}_{ia} =&
 \frac{\sqrt{3}}{8}
 \epsilon_{ikl}\overleftrightarrow{\nabla}_k
 (i\sigma_2\sigma_l)(i\tau_2\tau_a), \\
 P^{(^3P_2)}_{ija} =&
  \frac{\sqrt{3}}{8\sqrt{2}}
  \left[\overleftrightarrow{\nabla}_i(i\sigma_2\sigma_j)
  +\overleftrightarrow{\nabla}_j(i\sigma_2\sigma_i)
  -\frac{2}{3}\delta_{ij}
 \overleftrightarrow{\nabla}_k(i\sigma_2\sigma_k)\right](i\tau_2\tau_a),
 \\
 P^{(^3F_2)}_{ija}=&\frac{5}{32}
 \left[
 \overleftrightarrow{\nabla}_i \overleftrightarrow{\nabla}_j
 \overleftrightarrow{\nabla}_l
 -\frac{1}{3}
 \left(\overleftrightarrow{\nabla}_i
 \delta_{jl}+\overleftrightarrow{\nabla}_j
 \delta_{li}+\overleftrightarrow{\nabla}_l
 \delta_{ij}
 \right)
 \overleftrightarrow{\nabla}^2 
 \right]
 (i\sigma_2\sigma_l)(i\tau_2\tau_a)\, .                  
\end{align}
They are normalized in the same way as in Ref.~\cite{Fleming:1999ee},
\begin{equation}
 \sum_{\rm{pol. avg}}\Tr
  \left[ P^{(s)}P^{(s^{\prime})\dagger}\right]
  =\frac{1}{2}\left|\bfk\right|^{2l}\delta_{ss^{\prime}},
\end{equation}
where $\bfk$ is the relative three-momentum and $l$ is the orbital
angular momentum of the channel. For the $P$ waves $l=1$.  The operators
in another channel are obtained by replacing $P_i^{(^1P_1)}$ with the
corresponding projection operator. In the ${}^3P_2$-${}^3F_2$ channel,
there is an additional operator,
\begin{equation}
 C_4^{(^3F_2)}
  \left[
   \left(N^{T}P_{ija}^{(^3P_2)}N\right)^{\dagger}
   \left(N^{T}P_{ija}^{(^3F_2)}N\right)+h.c.
  \right],
\end{equation}
which represents the mixing.  Since the calculations are completely
parallel, we will only demonstrate the results in the $P_i^{(^1P_1)}$
channel in the following.


We introduce the dimensionless coupling constants,
\begin{equation}
 x_{(1,1)}= \frac{M\Lambda^3}{2\pi^2}C_2^{(^1P_1)},\quad 
 y_{(1,1)}= \frac{4M\Lambda^5}{2\pi^2}C_4^{(^1P_1)},\quad
 z_{(1,1)}= \frac{\Lambda^5}{2\pi^2}B_4^{(^1P_1)},
\end{equation}
to write the following RG equations in the $n\rightarrow\infty$ limit,
\begin{align}
 \frac{dx_{(1,1)}}{dt}=&-3x_{(1,1)}
 -\left[x_{(1,1)}^2+2x_{(1,1)}y_{(1,1)}+2x_{(1,1)}z_{(1,1)}+y_{(1,1)}^2
 +2y_{(1,1)}z_{(1,1)}+z_{(1,1)}^2 \right],\\
 \frac{dy_{(1,1)}}{dt}=&-5y_{(1,1)}
 -\left[\frac{1}{2}x_{(1,1)}^2+2x_{(1,1)}y_{(1,1)}+\frac{3}{2}y_{(1,1)}^2
 +y_{(1,1)}z_{(1,1)}-\frac{1}{2}z_{(1,1)}^2\right],\\
 \frac{dz_{(1,1)}}{dt}=&-5z_{(1,1)}
 +\left[\frac{1}{2}x_{(1,1)}^2+x_{(1,1)}y_{(1,1)}-x_{(1,1)}z_{(1,1)}
 +\frac{1}{2}y_{(1,1)}^2-y_{(1,1)}z_{(1,1)}-\frac{3}{2}z_{(1,1)}^2\right].
\end{align}
We obtain the RGEs of the same form for the ${}^3P_0$ and ${}^3P_1$
channels with the coupling constants being suitably defined.

Compare them with the RGEs for the ${}^1S_0$ channel, (\ref{xbeta}),
(\ref{ybeta}), and (\ref{zbeta}). With the appropriate replacement of
the coupling constants, the coefficients of the quadratic terms are the
same. Only the coefficients of the linear terms, which are nothing but
the canonical dimensions of the corresponding operators, are different.
This is because the channel dependence enters into the averaged action
only through the projection operator and the canonical dimensions of the
coupling constants. See the interaction part of Eq.~(\ref{second}) and
Eq.~(\ref{pwavegamma}), for example.

One can easily obtain the fixed points,
\begin{equation}
 (x^\star,y^\star,z^\star)=(0,0,0),\quad
  \left(-3,\frac{9}{2},-\frac{9}{2}\right),\quad
  \left(-\frac{25}{3},\frac{35}{6},-\frac{5}{2}\right).
\end{equation}
We find that the scaling dimensions at the third fixed point are complex
and may be disregarded. 

It is useful to define the following variables,
\begin{equation}
 u=x_{(1,1)},\quad v=\half \left(y_{(1,1)}-z_{(1,1)}\right), \quad
  w=\half  \left(y_{(1,1)}+z_{(1,1)}\right),
\end{equation}
in terms of which the RG equations are written as
\begin{align}
 \frac{du}{dt}&=-3u -(u+2w)^2 \\
 \frac{dv}{dt}&=-5v-\half(u+4v)(u+2w) \\
 \frac{dw}{dt}&=-5w-w(u+2w)
\end{align}
The nontrivial fixed point is now at $(u^\star, v^\star,
w^\star)=\left(-3, \frac{9}{2},0\right)$. Note that flows starting in
the $w=0$ plane never depart from it.

\begin{figure}
 \includegraphics[width=0.7\linewidth,clip]{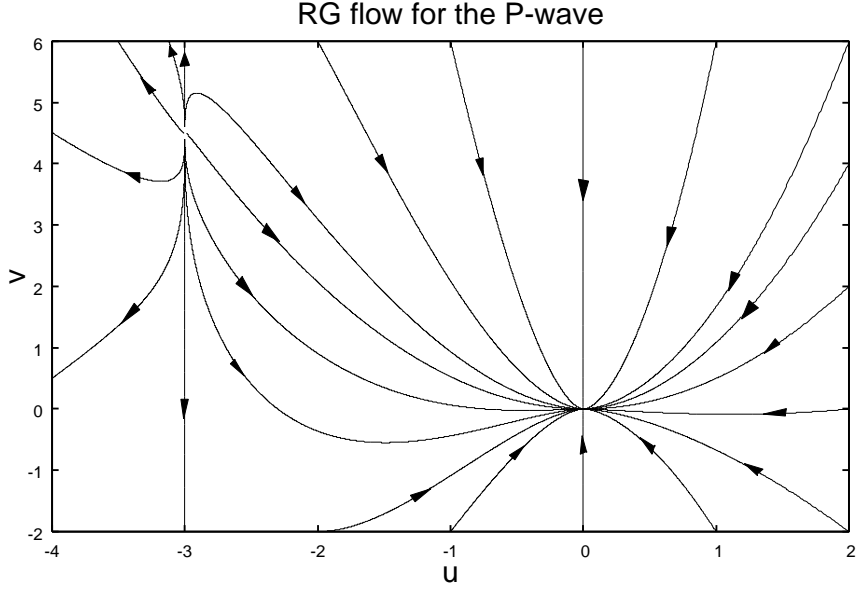}%
 \caption{\label{fig:pwave}The RG flow for the $P$ wave in the $w=0$ plane.}
 \end{figure}

In Fig.~\ref{fig:pwave}, we show the RG flow in the $w=0$ plane. It is
easy to see that $u=-3$ is a phase boundary; the flows in the right of
it (the weak coupling phase) run to the trivial fixed point, while those
in the left of it (the strong coupling phase) go to infinity.

At the nontrivial fixed point, we can easily obtain the following
scaling dimensions and corresponding eigenvectors (in the $u,v,w$
basis),
\begin{equation}
 \nu_1=3: \ \bfu_1=
 \left(
 \begin{array}{c}
  1\\
  -3 \\
  0
 \end{array}
 \right), \quad
\nu_2=1: \ \bfu_2=
 \left(
 \begin{array}{c}
  0\\
  1 \\
  0
 \end{array}
 \right), \quad
\nu_3=-2: \ \bfu_3=
 \left(
 \begin{array}{c}
  12\\
  -1 \\
  -5
 \end{array}
 \right). \label{p-eigen1}
\end{equation}
Note that there are two relevant directions, though they are all
irrelevant at the trivial fixed point. For the first two eigenvectors,
the scaling dimensions shift from their canonical values by six.


We can investigate the ${}^3P_2$-${}^3F_2$ channel in a similar
way. After introducing dimensionless couplings,
\begin{align}
 &x_{(3,2)}=\frac{M\Lambda^3}{2\pi^2}C_2^{(^3P_2)} , \quad
 y_{(3,2)}=\frac{4M\Lambda^5}{2\pi^2}C_4^{(^3P_2)},\nonumber \\
 &z_{(3,2)}=\frac{\Lambda^5}{2\pi^2}B_4^{(^3P_2)}, \quad 
 w_{(3,2)}=\frac{\sqrt{2}}{2}\frac{M\Lambda^5}{2\pi^2}C_4^{(^3F_2)},
\end{align}
we have (in the $n\rightarrow \infty$ limit)
\begin{align}
 \frac{dx_{(3,2)}}{dt}=&-\!3x_{(3,2)}
 \!-\!\left[x_{(3,2)}^2\!+2x_{(3,2)}y_{(3,2)}\!
 +2x_{(3,2)}z_{(3,2)}\!+y_{(3,2)}^2
 +2y_{(3,2)}z_{(3,2)}+z_{(3,2)}^2+2w_{(3,2)}^2 \right],\\
 \frac{dy_{(3,2)}}{dt}=&-5y_{(3,2)}
 -\left[\frac{1}{2}x_{(3,2)}^2+2x_{(3,2)}y_{(3,2)}+\frac{3}{2}y_{(3,2)}^2
 +y_{(3,2)}z_{(3,2)}-\frac{1}{2}z_{(3,2)}^2+w_{(3,2)}^2\right],\\
 \frac{dz_{(3,2)}}{dt}=&-\!5z_{(3,2)}
 \!+\!\left[\frac{1}{2}x_{(3,2)}^2\!+x_{(3,2)}y_{(3,2)}\!
 -x_{(3,2)}z_{(3,2)}
 \!+\frac{1}{2}y_{(3,2)}^2\!-y_{(3,2)}z_{(3,2)}
 \!-\frac{3}{2}z_{(3,2)}^2\!+w_{(3,2)}^2\right],\\
 \frac{dw_{(3,2)}}{dt}=&-5w_{(3,2)}-\left[
 x_{(3,2)}w_{(3,2)}+y_{(3,2)}w_{(3,2)}+z_{(3,2)}w_{(3,2)}\right].
\end{align}
Note that, with the appropriate replacement of the coupling constants,
the coefficients of the quadratic terms of the above RGEs are the same
as those in the ${}^3S_1$-${}^3D_1$ channel, and the only difference is
the coefficients of the linear terms representing the canonical
dimensions, just as we explained for the ${}^1P_1$ and ${}^1S_0$
channels. The flows and the scaling dimensions are similar to those for
other $P$ waves, just as those for the ${}^3S_1$-${}^3D_1$ channel to
those of ${}^1S_0$.

\subsection{Amplitude for the $P$ waves}
\label{sec:amplitude}

Similar results may be obtained by explicitly calculating the scattering
amplitude as we did for the $S$-waves in Refs.~\cite{Harada:2005tw,
Harada:2006cw}, which is the extention of the method in
Refs.~\cite{Phillips:1997xu, Gegelia:1998iu} to
include the redundant operators. Consider the Lippmann-Schwinger
equation with the ``potential'' in the center-of-mass frame (See
Appendix~\ref{sec:equiv}),
\begin{equation}
  -iV=-iP^\dagger\otimes P \left(-4\bfp_1\cdot\bfp_2\right) 
  \left[
   C_2
   +4C_4\left(\bfp_1^2+\bfp_2^2\right)
   -2B_4\left(p^0-\frac{\bfp_1^2+\bfp_2^2}{2M}\right)
  \right],
\end{equation}
where $\bfp_1$ and $\bfp_2$ are momenta of the nucleons in the initial
and final states respectively, $P$ stands for the spin-isospin factor of
the projection operator (for example, $P=\frac{\sqrt{3}}{4\sqrt{2}}
\sigma_2\tau_2$ for the $^1P_1$ channel), and the ansatz for the
amplitude,
\begin{equation}
 -i\mathcal{A}=-i P^\dagger\otimes P \left(-4\bfp_1\cdot\bfp_2\right) 
  \left[
   X(p^0)+Y(p^0)\left(\bfp_1^2+\bfp_2^2\right)+Z(p^0)\bfp_1^2 \bfp_2^2
  \right],
\end{equation}
where $X(p^0)$, $Y(p^0)$, and $Z(p^0)$ are functions to be
determined. The Lippmann-Schwinger equation reduces to
\begin{equation}
 -i\tilde{\mathcal{A}}(p^0,p_1,p_2)=-i\tilde{V}(p^0,p_1,p_2)
  +\int \!\!\frac{d^3k}{(2\pi)^3}\!
  \left(-i\tilde{V}(p^0, k,p_2)\right)
  \frac{ik^2}{p^0\!-\!\frac{k^2}{M}\!+\!i\epsilon}
  \left(-i\tilde{\mathcal{A}}(p^0,p_1,k)\right),
\end{equation}
where we have introduced
\begin{align}
 \tilde{\mathcal{A}}&=
 \left[
 X(p^0)+Y(p^0)\left(\bfp_1^2+\bfp_2^2\right)+Z(p^0)\bfp_1^2\bfp_2^2
 \right], \\
 \tilde{V}&=
 \left[
 C_2 +4C_4\left(\bfp_1^2+\bfp_2^2\right)
 -2B_4\left(p^0-\frac{\bfp_1^2+\bfp_2^2}{2M}\right)
 \right].
\end{align}
The solution is easily obtained as,
\begin{align}
 X&=\frac{1}{D}
 \left[
 \left(C_2-2B_4p^0\right)+\left(4C_4+\frac{B_4}{M}\right)^2I_3
 \right], \\
 Y&=\frac{1}{D}
 \left(4C_4+\frac{B_4}{M}\right)
 \left[
 1-\left(4C_4+\frac{B_4}{M}\right)I_2
 \right], \\
 Z&=\frac{1}{D}\left(4C_4+\frac{B_4}{M}\right)^2I_1,
\end{align}
with
\begin{align}
 D=&1-\left(C_2-2B_4p^0\right)I_1
 -2\left(4C_4+\frac{B_4}{M}\right)I_2 \nonumber \\
 &+\left(4C_4+\frac{B_4}{M}\right)^2I_2^2
 -\left(4C_4+\frac{B_4}{M}\right)^2I_1I_3,
\end{align}
where 
\begin{equation}
 I_n=-\frac{M}{2\pi^2}\int_0^\Lambda dk\frac{k^{2n+2}}{k^2+\mu^2},\quad
  \mu=\sqrt{-Mp^0-i\epsilon}.
\end{equation}

The RG equation may be obtained by requiring the amplitude to be
independent of the cutoff $\Lambda$. By introducing
\begin{equation}
 \mathcal{X}=1+\frac{2}{5}w, \
  \mathcal{Y}=u-\frac{4}{7}w^2, \
  \mathcal{Z}=2(v+w)+\frac{4}{5}w^2,
\end{equation}
we find the following RG equations,
\begin{align}
 \Lambda\frac{d\mathcal{X}}{d\Lambda}&=
 \frac{1}{\mathcal{X}}
 \left(\mathcal{X}-1\right)\left(5\mathcal{X}^2+\mathcal{Y}\right), \\
 \Lambda\frac{d\mathcal{Y}}{d\Lambda}&=
 \frac{\mathcal{Y}}{\mathcal{X}^2}
 \left(
 10\mathcal{X}^3-7\mathcal{X}^2 + 2\mathcal{X}\mathcal{Y}-\mathcal{Y}
 \right), \\
 \Lambda\frac{d\mathcal{Z}}{d\Lambda}&=
 \frac{1}{\mathcal{X}^2}
 \left(
 \mathcal{Y}^2-5\mathcal{X}^2\mathcal{Z}+10\mathcal{X}^3\mathcal{Z}
 +2\mathcal{X}\mathcal{Y}\mathcal{Z}
 \right),
\end{align}
which give rise to the following fixed points,
\begin{equation}
 \left(
  \mathcal{X}^\star,\mathcal{Y}^\star,\mathcal{Z}^\star
 \right)
 =(1,0,0), \quad (1, -3, 9).
\end{equation}
In the original variables, they are
\begin{equation}
 (u^\star, v^\star,w^\star)=(0,0,0), \quad \left(-3, \frac{9}{2},0\right),
\end{equation}
consistent with the previous analysis using the Legendre flow
equation. Note that there are no additional nontrivial fixed points.

At the nontrivial fixed point, we can obtain the scaling dimensions and
the corresponding eigenvectors (in the $u$, $v$, $w$ basis),
\begin{equation}
 \nu_1=3: \ \bfu_1=
 \left(
 \begin{array}{c}
  1\\
  -3 \\
  0
 \end{array}
 \right), \quad
\nu_2=1: \ \bfu_2=
 \left(
 \begin{array}{c}
  0\\
  1 \\
  0
 \end{array}
 \right), \quad
\nu_3=-2: \ \bfu_3=
 \left(
 \begin{array}{c}
  12\\
  -13 \\
  -5
 \end{array}
 \right).  \label{p-eigen2}
\end{equation}
Compare the results with (\ref{p-eigen1}). The scaling dimensions agree
and the only difference is the eigenvector for $\nu_3$. We have seen a
similar phenomenon for the $^1S_0$ channel, as we explained in
Sec.~\ref{sec:p2}.

By substituting
\begin{equation}
 \left(
\begin{array}{c}
 \delta u\\
 \delta v\\
 \delta w
\end{array}
\right)
 =a \bfu_1 \left(\frac{\Lambda_0}{\Lambda}\right)^3
 +b \bfu_2 \left(\frac{\Lambda_0}{\Lambda}\right)
 +c \bfu_3 \left(\frac{\Lambda}{\Lambda_0}\right)^2
\end{equation}
into the amplitude, we obtain the off-shell amplitude near the
nontrivial fixed point,
\begin{equation}
 \left.\tilde{\mathcal{A}}^{-1}(p^0, \bfp_1,\bfp_2)\right|_*
  =-\frac{M\Lambda_0^3}{2\pi^2}\left(\frac{a}{9}\right)
  +\frac{M\Lambda_0}{2\pi^2}\left(-\frac{2b}{9}\right)\mu^2
  +\cdots +\frac{M}{4\pi}\mu^3,
  \label{amp-ab}
\end{equation}
where the ellipsis stands for higher order terms in $a$, $b$, and
$c$. By comparing it with the effective range expansion,
\begin{equation}
 \tilde{\mathcal{A}}^{-1}=-\frac{M}{4\pi}
  \left[-\frac{1}{\alpha}+\frac{r}{2}p^2+ \cdots -ip^3\right],
\end{equation}
defined on the mass shell, one sees
\begin{equation}
 \alpha^{-1}=-\frac{2\Lambda_0^3}{9\pi}a,
  \quad
  r=-\frac{8\Lambda_0}{9\pi}b.
\end{equation}
Namely, the two relevant directions correspond to the ``scattering
length'' $\alpha$ and the ``effective range'' $r$. (These effective
range parameters have different dimensionality in the $P$ waves from
those in the $S$ waves.)  Note that the strong coupling phase ($a <0$)
corresponds to a positive ``scattering length,'' and the weak coupling
phase ($a>0$) corresponds to a negative one.

In Ref.~\cite{Bertulani:2002sz} it is noticed that there are two
couplings to be fine-tuned. They corresponds to our relevant parameters
given above. We have given the explanation for what they found from the
RG point of view.

In order to understand the difference between the strong and the weak
coupling phases, we take a closer look at the poles of the amplitude in
the effective range expansion. Note that we are in the vicinity of the
nontrivial fixed point, i.e., $|a| \ll 1$ and $|b| \ll 1$. At low
energies, the poles are the solutions of the cubic
equation~\cite{Bertulani:2002sz},
\begin{equation}
 -\frac{1}{\alpha}+\frac{r}{2}p^2-ip^3=0.
\end{equation}
Note that if we change the signs of $\alpha$ and $r$ simultaneously,
$\alpha \rightarrow -\alpha$ and $r \rightarrow -r$, the solutions are
given by $p \rightarrow p^*$. 

Let us first suppose that we are in the region $\alpha^{-1}<0$ but not
very close to the critical value $\left|\alpha^{-1}\right| = 0$, i.e.,
$\left|\alpha r^3\right| <54$. In this case, we have three solutions,
\begin{equation}
 \pm\frac{\sqrt{3}}{12}
  \left(
   \frac{\left|\alpha\right|^{\frac{1}{3}}r^2}{v}
   -\frac{v}{\left|\alpha\right|^{\frac{1}{3}}}
  \right)
  -\frac{i}{6}
  \left(
   r+\frac{\left|\alpha\right|^{\frac{1}{3}}r^2}{2v}
   +\frac{v}{2\left|\alpha\right|^{\frac{1}{3}}}
  \right), \quad
  -\frac{i}{6}
  \left(
   r-\frac{\left|\alpha\right|^{\frac{1}{3}}r^2}{v}
   -\frac{v}{\left|\alpha\right|^{\frac{1}{3}}}
  \right).
  \label{pwave-sol}
\end{equation}
where
\begin{equation}
 v\equiv 
  \left[
   108+\alpha r^3+108\sqrt{1+\frac{\alpha r^3}{54}}
  \right]^{\frac{1}{3}}
\end{equation}
is a positive number. These poles are in most cases unimportant in
low-energy scattering. Only when we take the limit $\alpha^{-1}
\rightarrow 0$ and $r\rightarrow 0$ keeping the condition $\left|\alpha
r^3\right| <54$, all of the three poles move closer to the origin and
they become significant. The very existence of such a special limit is a
consequence of the two relevant directions. Note that this limit is a
simultaneous limit $a\rightarrow 0$ and $b\rightarrow 0$ with keeping
$|b^3/a|$ ``small'', i.e., $|b^3/a|<2187 \pi^2/128\approx 168.5$.  We do
not consider such a case here, but approach to the critical value
$\alpha^{-1} \rightarrow 0-$ keeping $r$ finite. This is the limit
$a\rightarrow 0$ keeping $b$ finite so that $|b^3/a| \rightarrow
\infty$. The discussion about these limits and the relations to the
power counting suggested in Refs.~\cite{Bertulani:2002sz,
Bedaque:2003wa} is given in Sec.~\ref{sec:summary}.

\begin{figure}[tbp]
\begin{tabular}{cc}
\begin{minipage}[t]{0.45\linewidth}
\begin{flushleft}
\includegraphics[width=\linewidth,clip]{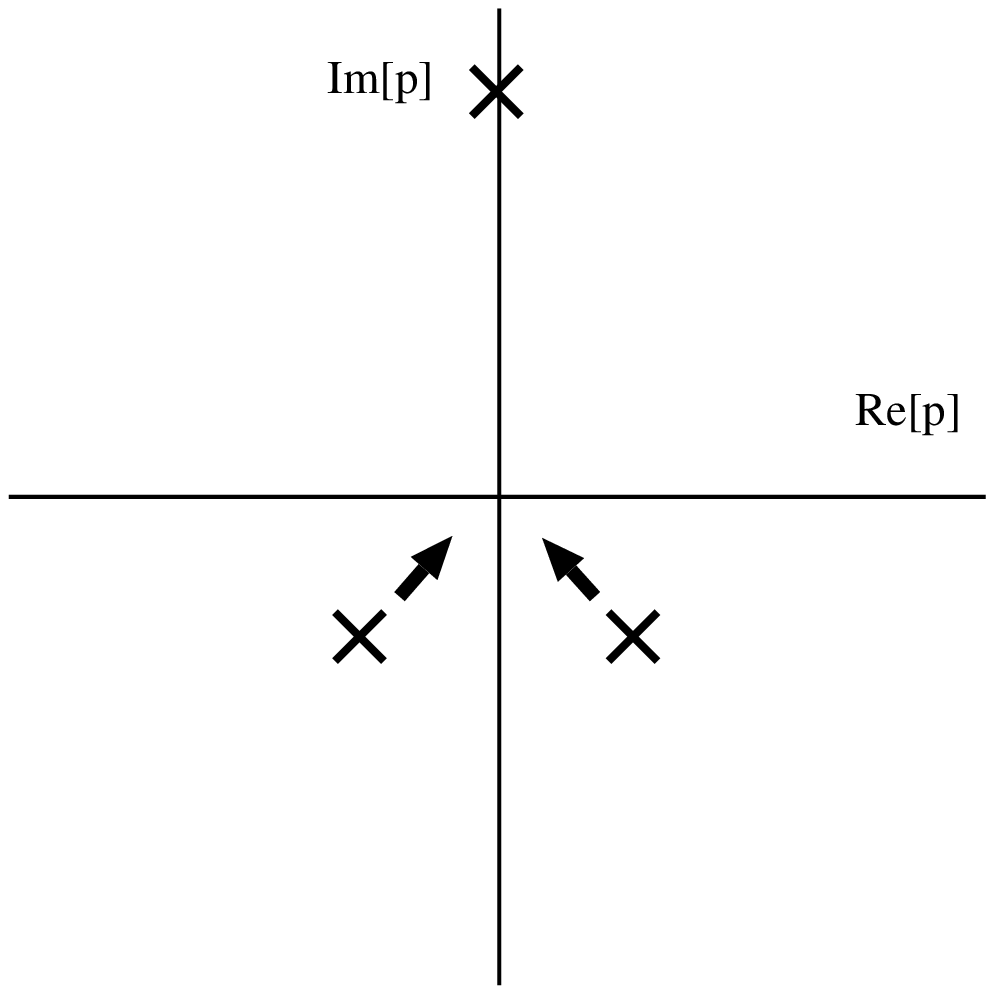}
\caption{\label{poles1} Poles for small $\alpha^{-1}<0$ and $r<0$. The arrows
 indicate the directions to which the poles move as $\alpha^{-1} \rightarrow
 0-$. There is a shallow resonance.}
\end{flushleft}
\end{minipage}\quad\quad\quad
\begin{minipage}[t]{0.45\linewidth}
\begin{flushright}
\includegraphics[width=\linewidth,clip]{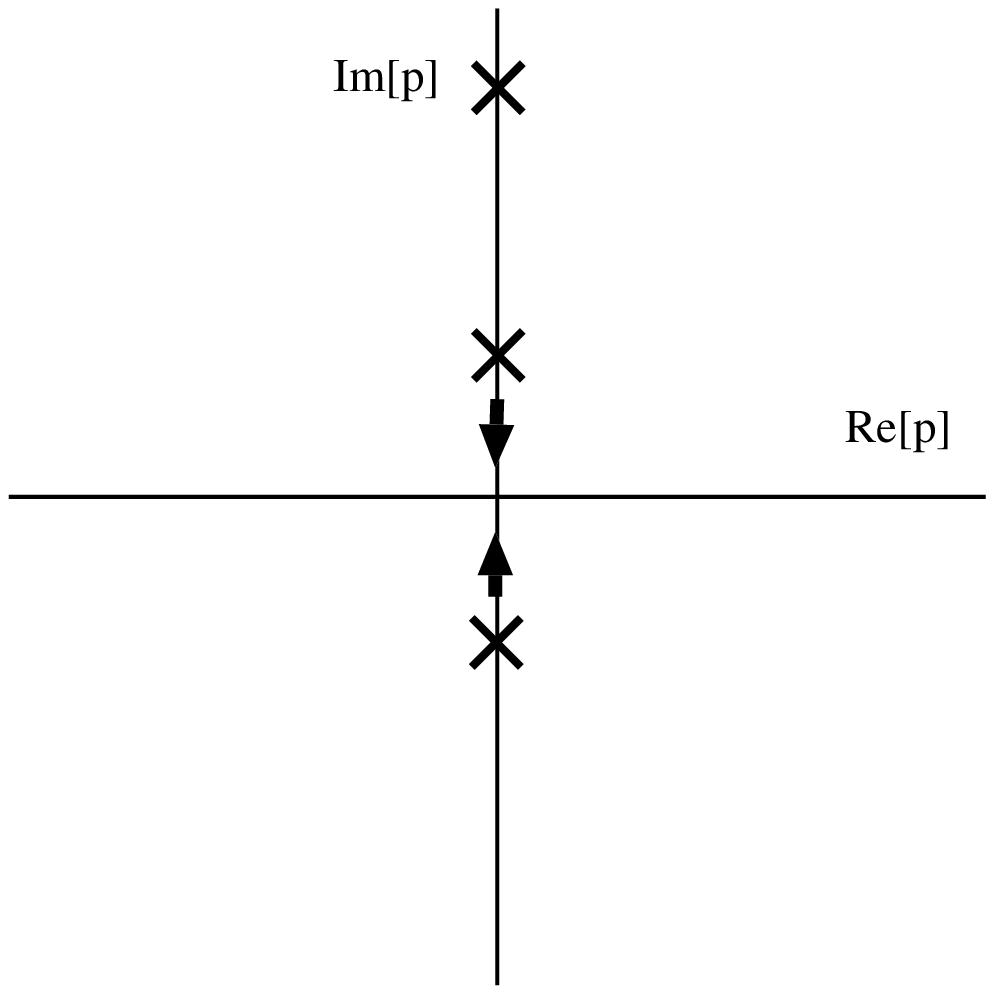}
 \caption{\label{poles2} Poles for small $\alpha^{-1}>0$ and $r<0$. The arrows
 indicate the directions to which the poles move as $\alpha^{-1} \rightarrow
 0+$. There is a shallow bound state.}
\end{flushright}
\end{minipage}
\end{tabular}
\end{figure}

If $r<0$, the expression for the solutions of Eq.~(\ref{pwave-sol}) is
still valid for $\left|\alpha r^3\right|>54$. The first two poles in
Eq.~(\ref{pwave-sol}) move toward the origin as we decrease the
magnitude of $\alpha^{-1}$ and reach it at $\alpha^{-1}=0$ (See
Fig.~\ref{poles1}), while the third one does not move so much. This
fairly deeply bound state is insignificant for low-energy
scattering. Note that the first two poles represent a shallow resonance.

If we go across the critical value $\alpha^{-1}=0$ into the region where
$\alpha^{-1}>0$, the two (degenerate) poles at the origin split on the
imaginary axis. We have the solutions
\begin{equation}
 -\frac{i}{6}r
  \left(
   1+2\sin\left(\frac{\phi}{3}\right)
  \right),
  \quad
  -\frac{i}{6}r
  \left(
   1\pm \sqrt{3}\cos\left(\frac{\phi}{3}\right)
   -\sin\left(\frac{\phi}{3}\right)
  \right),
  \label{pwave-strong}
\end{equation}
where
\begin{equation}
 \phi=\mbox{Arg}
  \left[
   \left(108 + \alpha r^3\right)i 
   +108 \sqrt{-\left(1+\frac{\alpha r^3}{54}\right)}
  \right].
\end{equation}
The first pole represents a shallow bound state. See Fig.~\ref{poles2}.
The pole with the positive sign of the second corresponds to the 
bound state found in the $\alpha^{-1}<0$ case. The pole with the
negative sign of the second does not seem to have a definite physical
meaning, but is responsible for the rapid change of the phase shift at
low momenta.

Note that there are several cases (the case with $r>0$ and $\alpha>0$
and the case with $r<0$, $\alpha>0$, and $|\alpha r^3|<54$) in which the
poles appear on the upper half plane, but causality prohibits poles on
the upper half plane except on the imaginary axis~\cite{PhysRev.83.249,
SchuetzerTiomno, PhysRev.91.1267, PhysRev.177.1848}. One should
therefore exclude such regions of $a$ and $b$ which give rise to acausal
poles.  See Refs.~\cite{Pham:1985cr, Ananthanarayan:1994hf,
Adams:2006sv, Distler:2006if, Manohar:2008tc, Adams:2008hp} for the
cases where some physical axioms such as causality, positivity and
unitarity constrain the values of low-energy constants. Remember that we
have assumed that $|a|\ll1$ and $|b|\ll1$ (and $|b^3/a|$ is large), that
is, $\left|\alpha^{-1}\right| \ll \Lambda_0^3$ and $|r| \ll \Lambda_0$,
with $|\alpha r^3|$ being large, through the analysis to meet the
condition that we are in the vicinity of the nontrivial fixed point. The
poles we have discussed have magnitudes smaller than the cutoff
$\Lambda_0$ to be consistent with the effective range expansion.

To summarize, we have done the RG analysis based on the cutoff
independence of the amplitude, and found that, near the nontrivial fixed
point, the scattering length and the effective range are related to the
deviations $a$ and $b$ from the nontrivial fixed point. The amplitude
has three poles at low momenta. Two of the three poles are sensitive
to the value of $\alpha^{-1}$ and get close to the origin near the
critical value. For $r<0$, there is a shallow resonance in the weak
coupling phase $a>0$ ($\alpha^{-1} <0$), and a shallow bound state in
the strong coupling phase $a<0$ ($\alpha^{-1} >0$). The appearance of a
(shallow) bound state in the strong coupling phase is similar to that in
the $S$ waves. But the appearance of a (shallow) resonance is a new
feature which is absent in the $S$ waves.

\subsection{$D$ waves}
\label{sec:dwave}

There are four channels in the $D$ waves; ${}^1D_2$, ${}^3D_1$,
${}^3D_2$, and ${}^3D_3$-${}^3G_3$, the second of which mixes with
${}^3S_1$ and has been considered. We introduce the following projection
operators,
\begin{align}
 P^{(^1D_2)}_{ija}=&\frac{\sqrt{15}}{16}
 \left(\overleftrightarrow{\nabla}_i\overleftrightarrow{\nabla}_j
 -\frac{1}{3}\overleftrightarrow{\nabla}^2\delta_{ij}\right)
 (i\sigma_2)(i\tau_2\tau_a) , \nonumber\\
 P^{(^3D_2)}_{ij}=&\frac{\sqrt{5}}{16\sqrt{2}}
 \left[
 \epsilon_{imn}\overleftrightarrow{\nabla}_m\overleftrightarrow{\nabla}_j
 +\epsilon_{jmn}\overleftrightarrow{\nabla}_m\overleftrightarrow{\nabla}_i
 \right]
 (i\sigma_2\sigma_n)(i\tau_2), \nonumber \\
 P^{(^3D_3)}_{ijk}=&
 \frac{\sqrt{15}}{48}\Bigg[
 \left\{
 \left(
 \overleftrightarrow{\nabla}_i
 \overleftrightarrow{\nabla}_j
 -\frac{1}{3}\overleftrightarrow{\nabla}^2\delta_{ij}
 \right)
 (i\sigma_2\sigma_k)
 + (\mbox{cyclic in $i,j,k$})
 \right\} \nonumber\\
&-\frac{2}{5}
 \left\{
 \left(
 \overleftrightarrow{\nabla}_k
 \overleftrightarrow{\nabla}_l
 -\frac{1}{3}\overleftrightarrow{\nabla}^2\delta_{kl}
 \right)\delta_{ij}
 +(\mbox{cyclic in $i,j,k$})
 \right\}
 (i\sigma_2\sigma_l)
 {\Bigg ]}(i\tau_2), \nonumber \\
 P^{(^3G_3)}_{ijk}=&
\frac{7\sqrt{5}}{128}\Bigg [
 \overleftrightarrow{\nabla}_i
 \overleftrightarrow{\nabla}_j
 \overleftrightarrow{\nabla}_k
 \overleftrightarrow{\nabla}_l
 -\frac{1}{7}\overleftrightarrow{\nabla}^2
 \left(
 \overleftrightarrow{\nabla}_i
 \overleftrightarrow{\nabla}_j
 \delta_{kl}
+(\mbox{$5$ terms to be symmetric in $i,j,k,l$})
 \right)\nonumber\\
 &+
 \frac{1}{35}\overleftrightarrow{\nabla}^4
 \left(
 \delta_{ij}\delta_{kl}
 +\delta_{ik}\delta_{jl}
 +\delta_{il}\delta_{jk}
 \right)
 \Bigg ](i\sigma_2\sigma_l)(i\tau_2).
\end{align}
Because the RG equations are the same for both ${}^1D_2$ and ${}^3D_2$
channels, we present the calculation for the ${}^1D_2$ channel
only. (Those for the coupled channel ${}^3D_3$-${}^3G_3$ are similar to
those for the ${}^3S_1$-${}^3D_1$ and the ${}^3P_2$-${}^3F_2$ channels
that we do not show them explicitly.) For this channel, the interaction
terms for the $\Gamma_\Lambda$ up to $\mathcal{O}(p^6)$ may be written
as
\begin{align}
 \Gamma_\Lambda^{int}=&\int d^4 x
 \Bigg[
 - C_4^{(^1D_2)}
 \left(N^{T}P_{ija}^{(^1D_2)}N\right)^{\dagger}
 \left(N^{T}P_{ija}^{(^1D_2)}N\right)
 \nonumber\\
 &+ C_6^{(^1D_2)}\left\{
 \left(N^{T}P_{ija}^{(^1D_2)}N\right)^{\dagger}
 \left(N^{T}P_{ija}^{(^1D_2)}\overleftrightarrow{\nabla}^2N\right)
 +h.c. \right\}\nonumber\\
 &+ 2B_6^{(^1D_2)}
 \left[
 \left(N^{T}P_{ija}^{(^1D_2)}N\right)^{\dagger}
 \left\{N^{T}P_{ija}^{(^1D_2)}
 \left(i\partial_{t}+\frac{\nabla^2}{2M}\right)
 N\right\}+ h.c.\right]
 \Bigg].
\end{align}
In terms of the dimensionless coupling constants,
\begin{equation}
 x_{(1,2)}= \frac{M\Lambda^5}{2\pi^2}C_4^{(^1D_2)}\, ,\quad 
 y_{(1,2)}= \frac{4M\Lambda^7}{2\pi^2}C_6^{(^1D_2)}\, ,\quad
 z_{(1,2)}= \frac{\Lambda^7}{2\pi^2}B_6^{(^1D_2)},
\end{equation}
we find the following RG equations (in the $n\rightarrow \infty$ limit),
\begin{align}
 \frac{dx_{(1,2)}}{dt}=&-5x_{(1,2)}
 -\left[x_{(1,2)}^2+2x_{(1,2)}y_{(1,2)}+2x_{(1,2)}z_{(1,2)}
 +y_{(1,2)}^2 +2y_{(1,2)}z_{(1,2)}+z_{(1,2)}^2 \right],\nonumber \\
 \frac{dy_{(1,2)}}{dt}=&-7y_{(1,2)}
 -\left[\frac{1}{2}x_{(1,2)}^2+2x_{(1,2)}y_{(1,2)}+\frac{3}{2}y_{(1,2)}^2
 +y_{(1,2)}z_{(1,2)}-\frac{1}{2}z_{(1,2)}^2\right],\nonumber\\
 \frac{dz_{(1,2)}}{dt}=&-7z_{(1,2)}
 +\left[
 \frac{1}{2}x_{(1,2)}^2+x_{(1,2)}y_{(1,2)}-x_{(1,2)}z_{(1,2)}
 +\frac{1}{2}y_{(1,2)}^2-y_{(1,2)}z_{(1,2)}-\frac{3}{2}z_{(1,2)}^2
 \right].
\end{align}
Just as for the ${}^1P_1$ case, with the appropriate replacement of the
coupling constants, the coefficients of the quadratic terms of the above
RGEs are the same as those in the ${}^1S_0$ channel, and the only
difference is the coefficients of the linear terms representing the
canonical dimensions. 

The nontrivial fixed points are found to be
\begin{equation}
 \left(x_{(1,2)}^\star,y_{(1,2)}^\star,z_{(1,2)}^\star\right)
  =(0,0,0),\quad
  \left(-5,\frac{25}{6},-\frac{25}{6}\right),  \quad
  \left(-\frac{49}{5},\frac{63}{10},-\frac{7}{2}\right),
\end{equation}
but the last nontrivial fixed point has complex scaling dimensions, and
thus may be disregarded. At the nontrivial fixed point, $
\left(-5,\frac{25}{6},-\frac{25}{6}\right)$, we find the following
scaling dimensions and the corresponding eigenvectors,
\begin{equation}
  \nu_1=5:\ u_1=
\left(
\begin{array}{c}
 3\\
 -5\\
 5\\
\end{array}
\right),
\quad
\nu_2=3:\ u_2=
\left(
\begin{array}{c}
 0\\
 -1\\
 1\\
\end{array}
\right),
\quad
\nu_3=-2:\ u_3=
\left(
\begin{array}{c}
 -10\\
 5\\
 2\\
\end{array}
\right).
\end{equation}
Note that there are two relevant couplings. For the first two
eigenvectors, the scaling dimensions shift from their canonical values
by ten.

To summarize, we have examined the $D$ waves in the similar way, and
found that there are two relevant couplings with the scaling dimensions
shifted by ten.  These large anomalous dimensions suggest that there may
be more relevant operators.  In the next section, we give another guide
for what we expect for scaling dimensions from a different point of
view, and show that there is one more relevant operator.

\subsection{PDS for higher partial waves}

In the previous paper~\cite{Harada:2006cw}, we explain how the PDS
leads to the ``shift by two'' rule for the $S$ waves, that is, the
anomalous dimensions of the four nucleon operators are two. In short,
the usual PDS renormalization, which subtracts the contribution at the
$D=3$ pole as well, treats the operators \textit{as if they were in
three dimensional spacetime}. The canonical dimensions of the
four-nucleon operators shift by two, e.g., the operator $(N^TN)^\dagger
(N^TN)$ has dimension six in $(1+3)$ dimensions, but four in $(1+2)$
dimensions. (In $D$ dimensions, it has dimension $2(D-1)$.)

Note that this ``shift rule'' does not apply to the redundant operators,
which do not need to be introduced with the dimensional regularization.

We have seen that the rule for the $P$ waves is ``shift by six,'' and
for the $D$ waves, ``shift by ten.'' It is now natural to extend the PDS
scheme to higher partial waves by subtracting the contribution at $D=1$
for the $P$ waves, and that at $D=-1$ for the $D$ waves. 

One can easily show that this generalization of PDS for higher partial
waves works well as for the $S$ waves in the pionless NEFT.

This kind of generalization of PDS has been considered in
Ref.~\cite{Phillips:1998uy} but apparently the relevance to higher
partial waves was not noticed.

Guided by this extended PDS prescription, we expect that there should be
one more relevant operator (with the scaling dimension one) in the $D$
waves, even though we found only two relevant operators in the previous
subsection.  The operator will be found if one performs a similar
analysis in larger operator space.

Note that the RGEs for the $P$ and $D$ waves have the similar structure
to those for the $S$ waves, as we remarked in Sec.~\ref{sec:pwave} and
in Sec.~\ref{sec:dwave}.  The RGEs for the $D$ waves to
$\mathcal{O}(p^8)$ should be obtained from those for the $S$ waves to
$\mathcal{O}(p^4)$ given in Sec.~\ref{sec:rge} by replacing the
coefficients of the linear terms with the corresponding ones for the $D$
waves (which are nothing but the canonical dimensions), with appropriate
replacement of the coupling constants. By examining the RGEs, one can
immediately obtain the nontrivial fixed point,
\begin{equation}
 \left(
  -5, \frac{25}{6},-\frac{25}{6},0,
  \frac{100}{9},0,-\frac{100}{9}, -\frac{100}{9}
 \right),
\end{equation}
where the first three components are $(x_{(1,2)}, y_{(1,2)},
z_{(1,2)})$, while the rest are the dimensionless coupling constants of
$\mathcal{O}(p^8)$. 

By solving the linearized RGEs around it, we obtain the following
eigenvalues and eigenvectors,
\begin{align}
 &\nu_1=+5: \quad \bfu_1=
 \left(
  \frac{3}{5}, -1, 1, 0, -1, 0, 1, 1
 \right), \\
 &\nu_2=+3: \quad \bfu_2=
 \left(
 0, \frac{3}{20}, -\frac{3}{20}, 0, -1, 0, 1, 1
 \right), \\
 &\nu_3=+1: \quad \bfu_3=
 \left(
 0, 0, 0, 0, -1, 0, 1, 1
 \right), \\
 &\nu_4=-2: \quad \bfu_4=
 \left(
 -\frac{54}{127}, \frac{90}{127}, -\frac{261}{635}, 0, -\frac{190}{127}, 
 0, \frac{64}{127}, 1
 \right), \\
 &\nu_5=-4: \quad \bfu_5=
 \left(
 -\frac{30}{37}, \frac{25}{37}, -\frac{25}{37}, \frac{27}{37}, -2, 0, 0, 1
 \right), \\
 &\nu_6=-4: \quad \bfu_6=
 \left(
 -\frac{45}{74}, -\frac{55}{74}, \frac{55}{74}, \frac{81}{148}, 1, 0, 1, 0
 \right), \\
 &\nu_7=-4: \quad \bfu_7=
 \left(
 -\frac{45}{37}, -\frac{55}{37}, \frac{55}{37}, \frac{155}{74}, 0, 1, 0, 0
 \right), \\
 &\nu_8=-9: \quad \bfu_8=
 \left(
 \frac{9}{8}, -\frac{15}{8}, \frac{3}{10}, 0, \frac{13}{8}, 0, \frac{71}{50}, 1
 \right).
\end{align}
We find the third relevant operator as we expected. 
The extended PDS prescription and the Wilsonian RG analysis given above
give a consistent result.

\section{Summary and Discussion}
\label{sec:summary}

In this paper, we extend our previous study of the Wilsonian RG analysis
for the pionless NEFT in the two-nucleon sector. The determination of
power counting based on the scaling dimensions is a simple and powerful
method in particular for the theories in which nonperturbative dynamics
is important.

Two kinds of extensions are considered; (1) enlargement of the space of
operators to be taken into account, and (2) higher partial
waves. Because our formulation is general, we can use the same machinery
to analyze them.

We considered the space of operators up to including those of
$\mathcal{O}(p^4)$ in the $^1S_0$ channel and found that the results are
stable against the enlargement. We also found that there are ``fixed
lines'' and a ``fixed surface'' which are related to the existence of
marginal operators.

In the $P$ and $D$ waves, we derived the RG equations and found the
phase structures. There are two phases, the strong coupling and the weak
coupling phases, just as in the $S$ waves. Unlike the case of the $S$
waves, however, there are two relevant directions at the nontrivial
fixed point for the $P$ waves.  We found that there are three for the
$D$ waves.

By explicitly calculating the off-shell amplitude for the $P$ waves, we
have seen that (near the critical surface and $r<0$) there is a shallow
bound state in the strong coupling phase, while in the weak coupling
phase there is a shallow resonance.

To summarize, we have a coherent picture of the pionless NEFT in the
two-nucleon sector from the Wilsonian RG point of view.

In the following, we discuss several aspects of the results.
\begin{enumerate}
 \item In the enlarged space calculation, we found ``fixed lines'' and a
       ``fixed surface.'' At first sight, they seem very strange. But
       actually, their existence is related to that of marginal
       operators. In relativistic field theory, classically marginal
       operators usually get (non-integer) anomalous dimensions and turn
       into relevant or irrelevant operators, so that the corresponding
       coupling constants run though slowly. In the present case,
       we have no logarithmic divergences so that the marginal operators
       are really marginal. Their couplings do not run at all.

 \item The absence of logarithmic divergences seems to lead to integer
       scaling dimensions. We found however that ``fixed surface'' [F]
       has irrational scaling dimensions. This fact tempts us to think
       that it is an artefact.

 \item The additional nontrivial fixed point found in
       Ref.~\cite{Harada:2006cw}, which has complex scaling dimensions,
       disappears in the enlarged space calculation.  It is well known
       that the truncation in general produces spurious
       solutions~\cite{Margaritis:1987hv, Morris:1999ba}. It is also
       known that truncation leads to complex scaling dimensions. (Note
       that the RG equations obtained from amplitudes do not have such
       spurious fixed points.) As we discussed in Sec.~\ref{sec:rge},
       those fixed points with complex scaling dimensions are considered
       as spurious. Note however that what we are doing is not the
       so-called \textit{polynomial expansion} (the expansion in powers
       of fields), but the \textit{derivative expansion} only. It is due
       to the Fermi-Dirac statistics and the nonrelativistic feature
       that there are only a finite number of interaction terms in the
       derivative expansion to a given order.  It is remarkable that the
       derivative expansion shows the similar symptom to that of the
       polynomial expansion seen in relativistic scalar theory.

 \item As we discussed in a previous paper~\cite{Harada:2006cw}, some of
       the eigenvectors seem to correspond to the directions in which
       the physical quantities remain unchanged. We suspect that the
       eigenvectors with scaling dimensions which do not obey the
       ``shift by two'' rule in the $S$ waves are such directions, and
       that there is a similar correspondence in each partial wave.

 \item In the $P$ wave calculation, we found that there are two relevant
       operators in a very natural way. The fact that there are two
       couplings to be fine-tuned has been noticed in
       Ref.~\cite{Bertulani:2002sz}.
       It requires a nonperturbative analysis to
       decide which couplings should be treated nonperturbatively and
       cannot be determined by perturbative consideration. In fact, they
       showed that two couplings should be fine-tuned by the explicit
       (nonperturbative) calculation using a dimeron.

 \item In Ref.~\cite{Bertulani:2002sz}, the case
       \begin{equation}
	\alpha^{-1}\sim M_{lo}^3, \quad r\sim M_{lo}, 
	 \quad \mbox{so that}\quad 
	 \left|\alpha r^3\right|\sim \mathcal{O}(1)
       \end{equation}
       is considered as an unnatural case, where $M_{lo}$ is a
       low-energy scale. In Ref.~\cite{Bedaque:2003wa}, on the other
       hand, another power counting is considered, in which only one
       combination of coupling constants is fine-tuned,
       \begin{equation}
	\alpha^{-1} \sim M_{lo}^2M_{hi},\quad r\sim M_{hi},
	 \quad \mbox{so that}\quad 
	 \left|\alpha r^3\right|\sim 
	 \mathcal{O}\left(\frac{M_{hi}^2}{M_{lo}^2}\right) \gg 1
       \end{equation}
       where $M_{hi}$ is a high-energy scale. The former case
       corresponds to the special limit mentioned in
       Sec.~\ref{sec:amplitude}, in which $\alpha^{-1}$ is sent to zero,
       keeping $\left|\alpha r^3\right|<54$. All of the three poles move
       closer to the origin, thus they become significant in low-energy
       scattering. The latter corresponds to the limit in which
       $\alpha^{-1}\rightarrow 0$ keeping $r$ finite, so that
       $\left|\alpha r^3\right| >54$. Only two poles (representing a
       resonance in the weak coupling phase) move closer to the origin,
       while the third is insignificant. In terms of the phase diagram,
       this corresponds to the flows very close to the phase boundary,
       but not to the fixed point itself.

 \item In the real world, the two-nucleon system in the $P$ waves does
       not exhibit any shallow resonances nor bound states, so that it
       is in the weak coupling phase. There may be systems which can be
       described by the same EFT at low energies. For example, Feshbach
       resonances of ultracold $^{40}$K~\cite{PhysRevLett.90.053201} or
       ${}^6$Li~\cite{zhang:030702,schunck-2004} may be interesting.
\end{enumerate}


\appendix
\section{Relation between the Legendre flow equation in the sharp
 cutoff limit and the RG equation by Birse et al.}
\label{sec:equiv}

\subsection{Feynman rules}

\begin{figure}[tb]
\begin{tabular}{cc}
\begin{minipage}{0.45\linewidth}
  \begin{center}
 \includegraphics[width=0.4\linewidth,clip]{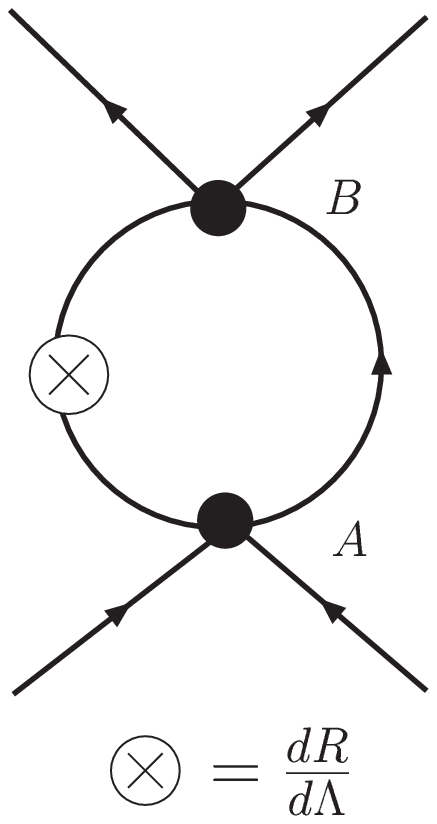}%
 \caption{\label{fig:AB-loop} One-loop diagram contributing to the
   Legendre flow equation. Labels $A$ and $B$ stand for the type of the
   vertices. The $A$ vertex is given in FIG.~\ref{fig:vertexA}.}
 \end{center}
\end{minipage}
\hspace{1cm}
\begin{minipage}{0.45\linewidth}
  \begin{center}
 \includegraphics[width=0.68\linewidth,clip]{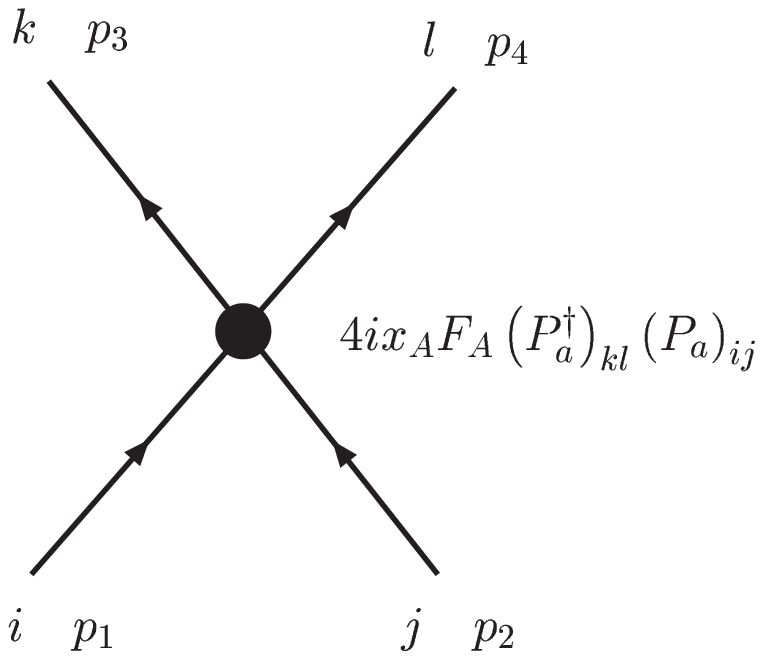}%
 \caption{\label{fig:vertexA} Feynman rule for a vertex labeled by $A$,
   where $i$, $j$, etc. are the spin-isospin indices, and $p_k$
   $(k=1,2,3,4)$ are momenta of the nucleons.}
 \end{center}
\end{minipage}
\end{tabular}
\end{figure}

Legendre flow equation reduces to a set of RG equations
for coupling constants, which consists of one-loop diagrams.  In the
present case, a typical diagram is given in Fig.~\ref{fig:AB-loop}. 
Let us first describe the Feynman rules for the ${}^1S_0$ channel as an
example. From the averaged action, one can easily read off the Feynman
rules for the vertex (See Fig.~\ref{fig:vertexA}),
\begin{equation}
 4i x_A F_A(p_f, p_i) \left(P_a^\dagger\right)_{kl}\left(P_a\right)_{ij},
\end{equation}
where $x_A$ is a dimensionless coupling constant. For the ${}^1S_0$
channel, it is one of the dimensionless coupling constants introduced in
(\ref{dimless}), $x_A\in \left\{x, y, z, u_1, u_2, z_1, z_2,
z_3\right\}$.  $\left(P_a\right)_{ij}$ is the spin-isospin factor
of the projection operator to the partial wave in question, with the
indices $i$ and $j$ referring to the spin and isospin quantum numbers of the
nucleon pair. 
$F_A(p_f, p_i)$ is the corresponding
momentum-dependent factor to $x_A$, where $p_i$ stands for the incoming
momenta, while $p_f$ for the outgoing ones. For the ${}^1S_0$ channel,
we have
\begin{align}
 F_x&=\frac{-2\pi^2}{M\Lambda}, \quad
 F_y=\frac{-2\pi^2}{4M\Lambda^3}\left(r_{12}+r_{34}\right), \quad
 F_z=\frac{2\pi^2}{\Lambda^3}\sum_{i=1}^4 S_i, \nonumber \\
 F_{u_1}&=\frac{-2\pi^2}{16M\Lambda^5}\left(r_{12}^2+r_{34}^2\right), 
 \quad
 F_{u_2}=\frac{-2\pi^2}{16M\Lambda^5}r_{12}r_{34}, \nonumber \\
 F_{z_1}&=\frac{2\pi^2M}{\Lambda^5}\sum_{i=1}^4 S_i^2, \quad
 F_{z_2}=\frac{2\pi^2M}{\Lambda^5}
 \left(S_1+S_2\right)\left(S_3+S_4\right), \nonumber \\
 F_{z_3}&=\frac{2\pi^2}{4\Lambda^5}
 \left\{
 r_{12}\left(S_3+S_4\right)+r_{34}\left(S_1+S_2\right)
 \right\}.
\end{align}
Note that $r_{ij}$ and $S_i$ are defined in Eq.~(\ref{def:r:S}).

\subsection{Subtleties at equal times}

When we include higher order redundant operators, we encounter
subtleties which do not emerge in the lowest order. In this subsection,
we explain the subtleties and how to handle them.

The Legendre flow equation may be obtained by considering the following
integral,
\begin{align}
 &\half
  \left(4ix_A\right)\left(4ix_B\right)
 \left(P_a^\dagger\right)_{kl}
 \Tr\left(P_aP_b^\dagger\right)
\left(P_b\right)_{ij}
 \nonumber \\
 &\times\int \frac{d^4k}{(2\pi)^4}
 F_A(p_i,k)
 \frac{i}
 {p_i^0/2+k^0-\mathcal{R}\left(\left(\bfp_i/2+\bfk\right)^2\right)+i\epsilon}
 \nonumber \\
 & \qquad \qquad \qquad \quad\times
 \frac{i}
 {p_i^0/2-k^0-\mathcal{R}\left(\left(\bfp_i/2-\bfk\right)^2\right)+i\epsilon}
 F_B(k,p_f),
 \label{integral}
\end{align}
where we have introduced 
\begin{equation}
 \mathcal{R}\left(\bfk^2\right)=
  \frac{\bfk^2}{2M}-R^{(1)}_{\Lambda}\left(\bfk^2\right),
\end{equation}
and $1/2$ is the symmetric factor. (Note that the way the IR cutoff
functions appear should be taken as symbolic, because it comes from the
``naive'' inclusion of the cutoff function in the averaged action that
however breaks Galilean invariance. A precise, Galilean invariant way is
given shortly.)

The problem is that the product of
the factors $F_A$ and $F_B$ may be quadratic or higher order in $k^0$,
so that the integral over $k^0$ appears to diverge. This divergence
cannot be regularized by the introduction of the higher order terms in
$k^0$ to the denominator of the propagator, because the results depend
on the inverses of (expectedly small) coefficients. It means that the
higher order terms drastically change the lower order results, and
cannot be accepted.

Note that in the relativistic field theory with dimensional
regularization such divergence does not cause a problem, because of
analytic continuation.

The divergence comes from the large $k^0$ region, in other words, from
the infinitely short time intervals. But from the EFT point of view, one
should have a cutoff on the energies of the intermediate states. (There
should be a finite resolution of time.) The cutoff may be of order
$\Lambda_0^2/M$, where $\Lambda_0$ is the physical cutoff on
three-momenta. We therefore assume, though implicitly, that $k^0$ is cutoff at
a scale of order $\Lambda_0^2/M$ and ignore the divergence arising from
the $k^0$ integration. Effectively, it ends up with evaluating the
integrand at the (either) pole. 

The manipulation of making the cutoff Galilean invariant is evident at
the amplitude level: to impose the cutoff on the relative
momentum~\cite{Harada:2006cw}. After doing so, Eq.~(\ref{integral})
becomes
\begin{equation}
 \half
  \left(4ix_A\right)\left(4ix_B\right)
  \left(P_a^\dagger\right)_{kl}
  \Tr\left(P_aP_b^\dagger\right)
  \left(P_b\right)_{ij}
  \int \frac{d^3k}{(2\pi)^3}
  \frac{i \left.F_A(p_i,k)F_B(k,p_f)\right|_{pole}}
  {A(p_i)-\bfk^2/M+2R_\Lambda^{(1)}(\bfk^2)+i\epsilon},
  \label{integral2}
\end{equation}
where we have introduced 
\begin{equation}
 A(P)\equiv P^0-\frac{\bfP^2}{4M},
\end{equation}
and $\left.F_A(p_i,k)F_B(k,p_f)\right|_{pole}$ denotes that
$F_A(p_i,k)F_B(k,p_f)$ is evaluated at the pole.

\subsection{Equivalence}

In order to establish the relation between the Legendre flow equation
and the RG equation employed by Birse et al., it is useful to rewrite
the Legendre flow equation in the sharp cutoff limit in a simpler form.

In the sharp cutoff limit, the integral in Eq.~(\ref{integral2}) may be
further simplified, and can be written as
\begin{equation}
 -4i x_A x_B
  \left(P_a^\dagger\right)_{kl} \left(P_a\right)_{ij}
  \frac{1}{8\pi^3}\int_\Lambda^\infty k^2dk 
  \int d^2\Omega_k
  \frac{\left.F_A(p_i,k)F_B(k,p_f)\right|_{pole}}
  {A(p_i)-k^2/M+i\epsilon},
\end{equation}
where we have used $\Tr\left(P_aP_b\right)=\delta_{ab}/2$. The change of
$\Lambda$ may be compensated by the change of the coupling
constants. Thus, if $C_C\sim x_C/\Lambda^{d_C}$, 
\begin{equation}
 \frac{dx_C}{dt}F_C(p_f,p_i) + d_C x_C F_C(p_f, p_i)
\end{equation}
receives the contribution from
\begin{align}
 &- x_A x_B
  \frac{1}{8\pi^3}\left(-\Lambda\frac{d}{d\Lambda}\right)
  \int_\Lambda^\infty k^2dk 
  \int d^2\Omega_k
  \frac{\left.F_A(p_i,k)F_B(k,p_f)\right|_{pole}}
  {A(p_i)-k^2/M+i\epsilon} \nonumber \\
 =& x_A x_B
  \frac{M\Lambda}{8\pi^3}
  \int d^2\Omega_k
  \frac{\left.F_A(p_i,\Lambda)F_B(\Lambda,p_f)\right|_{pole}}
  {1-\tilde{A}(p_i)},
\end{align}
where $\tilde{A}(P)=M A(P)/\Lambda^2$ and
$\left.F_A(p_i,\Lambda)F_B(\Lambda,p_f)\right|_{pole}$ is the simplified
expression for $\left.F_A(p_i,k)F_B(k,p_f)\right|_{pole}$ with the
magnitude of $k$ being $\Lambda$. In the center-of-mass frame, all the
$F_A$ in the ${}^1S_0$ channel do not depend on angles, so that we
obtain
\begin{equation}
 \frac{dx_C}{dt} + d_C x_C
  =\left.\sum_{A,B} x_A x_B
  \frac{M\Lambda}{2\pi^2}
  \frac{\left.F_A(p_i,\Lambda)F_B(\Lambda,p_f)\right|_{pole}}
  {1-\tilde{A}(p_i)}\right|_C,
  \label{sharpRGE}
\end{equation}
where $|_C$ stands for the operation of taking coefficient of
$F_C(p_f,p_i)$ in the expansion of the right hand side. This is a very
handy expression for the sharp cutoff limit.

If we identify the ``potential'' $V(p_f,p_i;\Lambda)$ as
\begin{equation}
  V(p_f, p_i; \Lambda) =-\sum_C x_C F_C(p_f, p_i),
\end{equation}
we have
\begin{align}
 \pd{V}{\Lambda}=& -\frac{1}{\Lambda}\sum_C
 \left(
 \Lambda\frac{dx_C}{d \Lambda}F_C-d_C x_C F_C
 \right) 
 = \frac{1}{\Lambda}\sum_C\left(\frac{dx_C}{dt}+d_C x_C\right)F_C.
\end{align}
Substituting Eq.~(\ref{sharpRGE}), we have
\begin{align}
 \pd{V}{\Lambda}=& \frac{M}{2\pi^2}\sum_{A}x_AF_A(p_f,\Lambda)
 \frac{1}{1-\tilde{A}(p_i)}\sum_{B}x_BF_B(\Lambda,p_i) \nonumber \\
 =& \frac{M}{2\pi^2} V(p_f,\Lambda; \Lambda)
 \frac{\Lambda^2}{\Lambda^2-MA(p_i)}V(\Lambda, p_i ; \Lambda).
\end{align}
Since $A(p_i)=p_i^0$ in the center-of-mass frame, it is nothing but the
RG equation (\ref{birse_rge}) employed by Birse et al.~\cite{Birse:1998dk}. 

\section{Cutoff function dependence of the results}
\label{sec:n-dep}

In this section, we briefly show how the results for the ${}^1S_0$
channel depend on the parameter $n$ in the cutoff function $R_\Lambda$
in Eq.~(\ref{cutoffR}).  We have derived RG equations for an arbitrary
value of $n$, but the expressions look too complicated that we omit
them. In the following, we refer the fixed points/lines/surface as [A],
[B], etc., as we defined in Sec.~\ref{sec:rge}.

The primary objective of this study is to justify the use of the sharp
cutoff limit, which gives the simplest expressions for the RG equations.
It is known that the sharp cutoff leads to bad behaviors 
in the derivative
expansion~\cite{Morris:1994ki}, so that it is important to
see how the results behave as we approach to the sharp cutoff limit.  In
particular, we are concerned with the possibility that the sharp cutoff
may cause non-analytic, or singular behaviors. We therefore consider the
smooth (i.e., differentiable) cutoffs which are very close to the sharp
cutoff limit. At any rate, we do not pretend to settle the convergence
problem caused by sharp cutoffs in the expansion. We only demonstrate
numerically that the use of sharp cutoff does not seem to cause a
serious problem in our present case.  On the other hand, some authors
try to extract the information about the convergence of the derivative
expansion by looking at the so-called ``scheme
(in-)dependence''~\cite{Litim:2002cf}. From our point of view, the
``scheme independence'' is a necessary, but not sufficient condition for
the convergence. The best we can do is to actually enlarge the space of
operators as we are doing in this paper.

The trivial fixed point [A] is always there, and has no $n$ dependence
at all. The $n$ dependence of the nontrivial fixed points [B] and [C] is
given in Tables~\ref{tab:b} and \ref{tab:c}. The fixed point [B] is very
stable against the variation of $n$. Even though the location changes
slightly, the scaling dimensions do not change at all. The fixed point
[C] has a stable limit but it has complex scaling dimensions. 

\begin{table*}
\caption{\label{tab:b}The dependence on $n$ of the fixed point [B].}
\begin{ruledtabular}
\begin{tabular}{ccc}
$n$&location of the fixed point&scaling dimensions\\
\hline 
$2$&{\scriptsize$(-1.10326, -0.604674, 0.604674, 0, 
 -1.67004, 0, 1.67004,1.67004)$}&$(-5,-4,-4,-4,-3,-2,-1,1)$\\
$10$&{\scriptsize$(-1.02722, -0.524991, 0.524991, 0,
 -1.40602, 0, 1.40602, 1.40602)$}&$(-5,-4,-4,-4,-3,-2,-1,1)$\\
$10^2$&{\scriptsize$(-1.00287, -0.502587, 0.502587, 0,
 -1.34072, 0, 1.34072, 1.34072)$}&$(-5,-4,-4,-4,-3,-2,-1,1)$\\
$10^3$&{\scriptsize$(-1.00029, -0.50026, 0.50026, 0,
 -1.33407, 0, 1.33407, 1.33407)$}&$(-5,-4,-4,-4,-3,-2,-1,1)$\\
$10^4$&{\scriptsize$(-1.00003, -0.500026, 0.500026, 0,
 -1.33341, 0, 1.33341, 1.33341)$}&$(-5,-4,-4,-4,-3,-2,-1,1)$\\
$\infty $&$(-1,-\frac{1}{2},\frac{1}{2},0,
 -\frac{4}{3},0,\frac{4}{3},\frac{4}{3})$&$(-5,-4,-4,-4,-3,-2,-1,1)$\\
\end{tabular}
\end{ruledtabular}
\end{table*}

\begin{table*}
\caption{\label{tab:c}The dependence on $n$ of the fixed point
 [C]. $\nu_i^2$ means that scaling dimension $\nu_i$ is doubly
 degenerate.}
\begin{ruledtabular}
\begin{tabular}{ccc}
$n$&location of the fixed point&scaling dimensions\\
\hline 
$2$&{\scriptsize$(-2.03159, 1.46753, 1.27909, 0., 
 -2.64366, 0., 1.78644, -1.84813)$}&
 {\scriptsize$(-6.36576, (-5.68288)^2, 5, 3,1,
 -1.68288 \pm 2.27988 i)$}\\
$10$&{\scriptsize$(-2.27317, 2.69046, 1.2095, 0.,
 -5.60121, 0., 1.3641, -2.62099)$}&
 {\scriptsize$(-7.85177, (-6.42588)^2, 5, 3,1,
 -2.42588 \pm 3.12036 i)$}\\
$10^2$&{\scriptsize$(-2.256, 2.79556, 0.984983, 0.,
 -5.98406, 0., 0.934766, -2.2433)$}&
 {\scriptsize$(-7.99819, (-6.49909)^2, 5, 3,1,
 -2.49909 \pm 3.20057 i)$}\\
$10^3$&{\scriptsize$(-2.25064, 2.79559, 0.957635, 0.,
 -5.99646, 0., 0.888947, -2.19038)$}&
 {\scriptsize$(-7.99998, (-6.49999)^2, 5, 3,1,
 -2.49999 \pm 3.20155 i)$}\\
$10^4$&{\scriptsize$(-2.25006, 2.79547, 0.954855, 0.,
 -5.99738, 0., 0.884356, -2.18494)$}&
 {\scriptsize$(-8, (-6.5)^2, 5, 3,1,
 -2.5 \pm 3.20156 i)$}\\
$\infty $&$(-\frac{9}{4},\frac{123}{44},\frac{21}{22},0,
 -\frac{92889}{15488},0,\frac{13689}{15488},-\frac{33831}{15488})$&
 $(-8, (-13/2)^2, 5, 3,1, (-5\pm i\sqrt{41})/2)$\\
\end{tabular}
\end{ruledtabular}
\end{table*}

We are not completely sure that the ``fixed lines'' and the ``fixed
surface'' exist for an arbitrary $n$, because we are unable to obtain
the analytic expressions for them. Numerical study indicates that at
least ``fixed line'' [E] is stable. In Table~\ref{tab:e}, we present the
$n$ dependence of a particular point ($u_2=0$) on the line and the
scaling dimensions at the point.

\begin{table*}
\caption{\label{tab:e}The dependence on $n$ of the fixed point $u_2=0$
 on the line [E]. $\nu_i^3$ means that scaling dimension $\nu_i$ is
 triply degenerate.}
\begin{ruledtabular}
\begin{tabular}{ccc}
$n$&location of the fixed point&scaling dimensions\\
\hline 
 $2$&{$(0, 0, 0, 0, 0, 0, 4.41305, 0)$}&
 {$((-5)^3,-3,-1,0,2,5)$}\\
$10$&{$(0, 0, 0, 0, 0, 0, 5.51631, 0)$}&
 {$((-5)^3,-3,-1,0,2,5)$}\\
$10^2$&{$(0, 0, 0, 0, 0, 0, 5.07010,0)$}&
 {$((-5)^3,-3,-1,0,2,5)$}\\
$10^3$&{$(0, 0, 0, 0, 0, 0, 5.00719, 0)$}&
 {$((-5)^3,-3,-1,0,2,5)$}\\
$10^4$&{$(0, 0, 0, 0, 0, 0, 5.00072, 0)$}&
 {$((-5)^3,-3,-1,0,2,5)$}\\
$\infty $&$(0,0,0,0,0,0,5,0)$&
 $((-5)^3, -3, -1, 0, 2, 5)$\\
\end{tabular}
\end{ruledtabular}
\end{table*}

\begin{acknowledgments}
 We would like to thank M.~C.~Birse for reading the manuscript and the
 very insightful discussions during his stay at Kyushu University.
 One of the authors (K.H.) is partially supported by Grant-in-Aid for
 Scientific Research on Priority Area, Number of Area 763, ``Dynamics of
 Strings and Fields,'' from the Ministry of Education, Culture, Sports,
 Science and Technology, Japan. 
\end{acknowledgments}

\bibliography{NEFT,NPRG,POTSCAT}

\end{document}